\documentclass[11pt,a4paper, aps, nofootinbib,onecolumn]{revtex4}

\usepackage{amsmath,amssymb,amsthm,amstext,amsfonts}
\usepackage[a4paper]{geometry}
\usepackage{makeidx}
\usepackage[pdftex]{graphics}
\usepackage{graphicx}
\usepackage{bbold}
\usepackage[english]{babel}
\usepackage{bm}
\usepackage[utf8]{inputenc}  
\usepackage[T1]{fontenc}       
\numberwithin{equation}{section}

\usepackage{helvet} 
\makeindex

\usepackage{graphics,graphicx,type1cm,eso-pic,color}

\makeatother

\renewcommand{\epsilon}{\varepsilon}
\renewcommand{\phi}{\varphi}
\newcommand\independent{\protect\mathpalette{\protect\independenT}{\perp}}
\def\independenT#1#2{\mathrel{\rlap{$#1#2$}\mkern2mu{#1#2}}}

\newtheorem{theorem}{Theorem}[section]

\newtheorem{proposition}[theorem]{Proposition}
\newtheorem{corollary}[theorem]{Corollary}
\newtheorem{definition}{Definition}
\newtheorem{assumption}{Assumption}

\newcommand{\E}{\mathbb{E}}
\newcommand{\beq}{\begin{equation}}
\newcommand{\eeq}{\end{equation}}

\usepackage{textcomp}

\begin{document}

\title{Copula-based Hierarchical Aggregation of Correlated Risks. \\The behaviour of the diversification benefit in Gaussian and Lognormal Trees}
\author{J-P. Bruneton}
\email{jean-philippe.bruneton@fundp.ac.be}
\affiliation{University of Namur, Department of Mathematics, 8 Rempart de la Vierge, B-5000 Namur, Belgium}
\date{\today}
\begin{abstract}
The benefits of diversifying risks are difficult to estimate quantitatively because of the uncertainties in the dependence structure between the risks. Also, the modelling of multidimensional dependencies is a non-trivial task. This paper focuses on one such technique for portfolio aggregation, namely the aggregation of risks within trees, where dependencies are set at each step of the aggregation with the help of some copulas. We define rigorously this procedure and then study extensively the Gaussian Tree of quite arbitrary size and shape, where  individual risks are normal, and where the Gaussian copula is used. We derive exact analytical results for the diversification benefit of the Gaussian tree as a function of its shape and of the dependency parameters. 

Such a ``toy-model'' of an aggregation tree enables one to understand the basic phenomena's at play while aggregating risks in this way. In particular, it is shown that, for a fixed number of individual risks, ``thin'' trees diversify better than ``fat'' trees. Related to this, it is shown that hierarchical trees have the natural tendency to lower the overall dependency with respect to the dependency parameter chosen at each step of the aggregation. We also show that these results hold in more general cases outside the gaussian world, and apply notably to more realistic portfolios (LogNormal trees). We believe that any insurer or reinsurer using such a tool should be aware of these systematic effects, and that this awareness should strongly call for designing trees that adequately fit the business.

We finally address the issue of specifying the full joint distribution between the risks. We show that the hierarchical mechanism does not require nor specify the joint distribution, but that the latter can be determined exactly (in the Gaussian case) by adding conditional independence hypotheses between the risks and their sums.
\end{abstract}

\maketitle
\newpage
\tableofcontents
\newpage
\section{Introduction}

In this introduction, we first review in Sections \ref{IntroDiversification} and \ref{IntroAggregation} some basics about diversification benefit, dependencies, risk measures and portfolio aggregation and the interplay between these notions from a Quantitative Risk Management perspective. In Sections~\ref{IntroTree} and \ref{IntroTreeII} we then present hierarchical aggregation trees and finally discuss the plan of the paper in Section~\ref{IntroPlan}.

\subsection{Diversification benefit and dependencies}
\label{IntroDiversification}
The core business of insurers and reinsurers is taking risks. Their ability to survive and to grow therefore critically depends on their capacity to diversify these risks, both on the liability and on the asset sides. Such a commonplace sadly gets a new flavor in these days as the financial industry is somehow discovering that some assets which have been considered for a long time as being risk-free --government bonds-- might not, after all, be so secure investments. 

Although the present paper does not deal with this critical issue, we believe that this very uncomfortable revolution for financial industry calls for having at hand robust and well-understood risk management tools. In these rather uncertain days, we think it is important to understand better how, and to which quantitative extent, financial industries benefit from the diversification of their portfolios.

The benefit of diversification decreases for correlated risks. The question of modelling accurately the diversification thus amounts to model adequately the dependencies between the risks. This is a very challenging task. Indeed, although the individual risks constituting the portfolio might easily be described with an appropriate stochastic model derived from data and/or expert opinion, it is often the case that very few joint observations are available, in which case the joint distribution of the individual risks is basically unknown. In this respect, the risks driven by extreme events are particularly relevant\footnote{In particular to reinsurance companies which underwrite excess of loss contracts.}, because then both marginal and joint observations are scarce \cite{EVT,Probex}. Moreover extreme risks have the tendency to correlate in the tail for which joint observations become extremely rare \cite{Donelly}. 

As a consequence, it is in general non-trivial to get a picture of the dependency structure between the individual risks. In the recent years, copulas \cite{Nelsen, EMBR} have become the privileged tool to overcome this difficulty. In a word, copulas are multivariate functions that allow separating the dependence structure from the margins. Copulas can then be used to define a dependence structure between margins, taking into account, in particular, the tail-dependencies. Moreover, there exist asymmetric copulas (such as the Clayton copula) which can reflect the asymmetry in the dependence between actual risks (risks which are correlated only in the tail) \cite{BurgiDacorogna}.

The present paper deals with the construction of a dependency structure based on copulas, via a hierarchical aggregation of risks. The diversification then becomes a function of the copulas used and also depends on the details of the hierarchical aggregation scheme. 

\subsection{Diversification, risk measures and portfolio aggregation}
\label{IntroAggregation}
In order to compute the diversification benefit, it is necessary to have a measure of the risk carried by the full portfolio: \textit{the sum at risk}. Then, the diversification benefit is measured as the ratio between the sum at risk of the actual portfolio to a fictive sum at risk in the case where the risks are fully dependent from each other (for a precise definition see Section~\ref{SecDB}). 

A widely used sum at risk is the Value-At-Risk ($\textrm{VaR}$) at some threshold $\alpha$, typically $\alpha =0.01$. More generally, the sum at risk is defined through a risk measure \cite{EMBR}. In this paper we will rather use the \textit{expected shortfall} at some threshold $\alpha$, $\textrm{ES}_{\alpha}$, also called the Tail-Value-At-Risk or $\textrm{TVaR}$. This risk measure has been shown to be more satisfying than the $\textrm{VaR}$ from a mathematical perspective, see  \cite{Artzner, EMBR}.

Because risk measures are not linear in their arguments however, $\textrm{TVaR}(\Sigma X_i) \neq \Sigma \,\textrm{TVaR} (X_i)$, it is necessary to know the distribution of the  total portfolio $Z= \sum X_i$ in order to compute the diversification benefit. This is what \textit{portfolio aggregation}, or \textit{risk aggregation} refers to: Portfolio aggregation is the computation of the distribution of $Z=\Sigma X_i$, given the marginals $X_i$ and a dependence structure (more on this in Section~\ref{SectionAggregationTree}). 

Although we focus on diversification in this paper, note that there are other motivations than pure risk management for aggregating risks together, such as e.g. capital management, capital allocation, business steering, and profitability analysis \cite{BurgiDacorogna}.
 
\subsection{Hierarchical aggregation of correlated risks}
\label{IntroTree}
Since one has only a limited knowledge about the dependency between the risks, the usual realistic framework for risk aggregation is to consider that only the distributions of the individual risks are available, but nothing or little on the joint distribution is known. Then one might aggregate the risks while assuming independence between them. This would obviously lead to a very poor estimate for the distribution of $Z$ and to a large overestimation of the diversification benefit. One might instead try to quantify the uncertainty on the sum at risk originating from the lack of information, via, e.g. computation of bounds on the sum at risk. Recently a general framework which interpolates between marginal knowledge and full knowledge of the joint distribution has been introduced along these lines in \cite{EmbrechtsRAgg}. More generally, the question of aggregating correlated risks has attracted lot of attention in the past years (for a short review see e.g. \cite{Wang}).

In the present paper we follow yet another route, where we assume enough information to be able to compute the total portfolio and the sum at risk (and hence the diversification) but without assuming that all the joint distribution between the individual risks is known. Moreover, this information is meant to be information actually attainable through actuarial work and expert opinion.

The basic idea is a \textit{hierarchical aggregation of risks}, which can be represented graphically as a \textit{tree of aggregation} (see Section~\ref{SectionAggregationTree}). To our knowledge, the idea was first presented in \cite[Chp.8]{Bluebook} and \cite{BurgiDacorogna}. The present article extends these works in many respects, see Sections~\ref{IntroTreeII} and \ref{IntroPlan}.

The idea is a step-by-step aggregation, where one first aggregates together risks within $N'$ disjoint subsets of the set of all the $N$ individual risks, by noticing that risks in the same subset share some common features (e.g. lines of business, region of the world, etc.). It makes then sense from an actuarial point of view to tie these risks together, i.e. to assume some dependency between these risks. The dependency is set via some suitably chosen and calibrated copula \cite{BurgiDacorogna,Probex}, which specifies the joint distribution between these risks. As a consequence, the respective $N'$ partial sums of the individual risks are then calculable. 

As a result of this first aggregation step, one ends up with partial sums of the individual risks. The procedure can then be repeated at upper levels, by noticing some other common features or dependencies between these partial sums, and tying them together via another set of copulas.  The process will be illustrated in Section~\ref{IntroTreeII} below and defined more rigorously in Section~\ref{SectionAggregationTree}. At the end of the process, one gets a random variable representing the total portfolio, the distribution function of which is not only a function of the individual risks, but also of the dependencies that have been chosen. 

When compared to the naive method of taking into account dependencies via a correlation matrix between the individual risks, the above hierarchical aggregation has several major advantages. First, it is based on copulas which are well-known to be more realistic measure of dependence (especially in the tail) than simple linear correlations. Second, and more importantly, the estimation of linear correlation coefficients is impossible in practice for large and realistic portfolios, because the number of unknowns goes as $N(N-1)/2$ for $N$ individual risks, whereas $N$ can be as large as $\sim 1000$ for worldwide companies. In an hierarchical tree of aggregation however, the number of parameters can be considerably reduced (see next sections), and it becomes feasible to calibrate them in a realistic way \cite{BurgiDacorogna,Probex}.

Note that one could have also chosen a direct copula-based aggregation (i.e. a flat tree, see Section~\ref{SectionAggregationTree}), by imposing directly some explicit $N-$dimensional copula between the individual risks. This would specify fully the joint distribution and allow in particular computing the total portfolio. As it is well-known however, it is very difficult to construct multivariate copulas, and only some families of them are known \cite{Nelsen,EMBR}. In particular, one would not be able to find (or even look for) a suitably defined $N-$dimensional copula that would reflect adequately the actual dependencies between all the individual risks. Indeed, it is important to realize that within all the individual risks, there might be both very dependent risks (e.g. wind France and wind Germany) but also uncorrelated risks (e.g. wind France and medical malpractice in Japan). Defining a $N-$dimensional copula given a set of such requirements would largely go beyond the current knowledge about copulas.

\subsection{Aggregation trees and their topology}
\label{IntroTreeII}
We saw that the use of hierarchical trees for portfolio aggregation has many advantages with respect to other methods. In short, it enables a construction of a viable dependency structure between the individual risks. Such a construction is realistic in the sense that it is based on copulas and requires only few information or parameters. It therefore solves (in principle) the difficult problem of risk aggregation, whereas its application is doable in practice (and indeed used by some companies). 
 
The great reduction of free parameters that we discussed should however not be misunderstood. It would indeed be misleading to think that the only freedom left in defining an aggregation tree lies in the calibration of the dependencies and the type of copulas used. In fact a lot of freedom has simply been hidden in the \textit{order} in which the risks are aggregated together and in the general \textit{shape} of the tree. Loosely speaking thus, the ``topology'' of the tree has a direct impact on the final distribution function, and therefore on the sum at risk carried by the full portfolio. 

The present paper will essentially focus on the role of the shape of the tree, not on the role of the order of aggregation. Let us however simply illustrate the latter on a concrete example. Consider four assets, two government bonds in countries 1 and 2, and respectively two stock market portfolios in countries 1 and 2. 

One might then consider the following aggregation tree:
\begin{figure}[ht]
\begin{center}
\includegraphics[width=8cm]{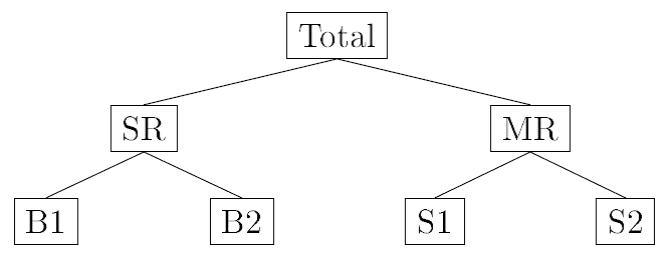}
\end{center}
\end{figure} 

Here SR stands for ``Sovereign Debt Risk'' (by definition, given by the sum of $B1$ and $B2$) and MR stands for ``Market Risk'' ($MR=S1+S2$). Also B1 stands for ``Bonds country 1'', and S1 for ``Stocks country 1''. It is clear however that one might also consider a different tree:
\begin{figure}[ht]
\begin{center}
\includegraphics[width=8cm]{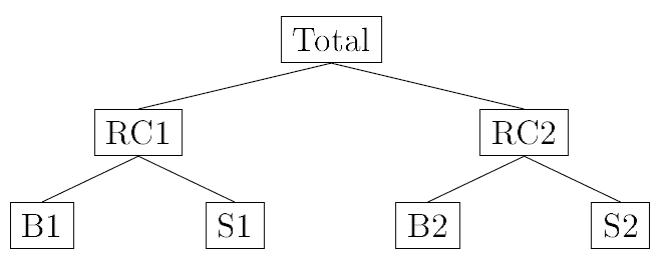}
\end{center}
\end{figure} 

where now RC1 stands for ``Risk of Country 1''. The difference between these two trees is simply the permutation ($B2 \leftrightarrow S1$), and possibly also the copulas used. In other words, the order of aggregation is different. Although these two trees aggregate the same individual risks with given marginals, they will not lead to the same distribution function for the total portfolio in general, except in trivial cases, e.g. if the identity copula is used. In particular, the sum at risk, and therefore the computed diversification benefit will differ.
  
This is both an advantage and a drawback of the method. It is a drawback in the sense that, even in such a simple case, an intrinsic ambiguity of the method arises. It is an advantage however, because this freedom in aggregating together the risks can be settled by actuarial experts with the aim of staying as close as possible to the business in the modelling choices. In the above fictive example, the choice between the two trees largely depends on whether the two countries 1 and 2 have strongly dependent economy (e.g. France and Germany) in which case the first tree is relevant or, in the opposite, if they have almost independent economy (then the second tree is relevant, because it models first regional risks, that are then connected via the last aggregation step, i.e. via the dependence on worldwide economy).

One of the conclusion is therefore that, as the world and the portfolios evolve with time, one should not only focus on regularly updating the dependencies parameters, but shall also be concerned with updating the structure of the tree itself and its adequacy with the risk profile of the company.

\subsection{Plan of the paper}
\label{IntroPlan}
Summarizing this introduction, we wish to address in this paper the issue of the aggregation of correlated risks and the impact of the modelling choices on the diversification benefit. There are various methods to do so. The present paper focuses on a particular method referred to as the hierarchical aggregation of risks, following the ideas found in \cite{BurgiDacorogna, Bluebook}.

The aim of this paper is twofold. First, we wish to provide a clean definition of the aggregation process described in this introduction, explain the rationale behind the graphical (tree) representation, and investigate the links with portfolio aggregation and incomplete information of the full risk profile (i.e. of the joint distribution). This will be done in Section~\ref{SectionAggregationTree} (see also Section~\ref{SecCI}).

Second, the main motivation is to study the influence of the  structure --or shape--  of the tree on the diversification benefit. As far as we are aware, this has not been studied before. We focus first on Gaussian Trees (to be defined in Section~\ref{SecGauss}). To get rid of the non-commutativity effect illustrated in Section~\ref{IntroTreeII}, we modeled every individual risks by the same distribution (a Gaussian), and we moreover applied the same (Gaussian) copula at every aggregation steps. Such a tree, that we dubbed ``Regular Gaussian Tree'' is admittedly of little phenomenological relevance because insurance risks usually have more fat tail distributions than the Gaussian. Also, realistic trees have no regular shape in general, and finally the Gaussian copula does not bring enough tail dependency between risks (see \cite{Davide}).

However, the Gaussian Tree has the major advantage of being completely solvable, and provide exact analytical -and very enlightening- results for the behaviour of the diversification benefit as a function of the shape of the tree (see Section~\ref{SecGauss}). One of the main conclusion is that ``thin'' trees diversify better than ``fat'' trees, meaning that the width and the depth of the tree is quite of importance regarding the diversification benefit, for a given fixed number of individual risks aggregated together. 

In particular, the results on the Gaussian tree naturally explains why one might still get quite an high diversification benefit although the dependencies set at each aggregation steps are high, provided the tree is thin enough. The explanation is rather intuitive, and amounts to the following: two individual risks which are far away from each other in the tree are connected via a node at the top of the tree, and therefore via the application of several times the Gaussian copula. This has the effect of lowering (possibly drastically) their effective coupling, ie. their correlation coefficient. This again makes sense from a business perspective (provided the tree is constructed in such a way that it fits the business), as there are no reasons why very different line of business like, say, Japan Earthquake and Medical Malpractice in UK should be anyhow connected to each other.
 
In order to go beyond the Gaussian model, we also provide similar numerical results for the LogNormal Trees and for various choices of copulas (see Section~\ref{SecLN}). We show that the behaviour seen in the Gaussian tree still holds in more realistic and more complex situations for which an analytical treatment is out of reach. In other words, and anticipating one of our conclusions, the Gaussian tree imposes itself as the natural benchmark to which any other trees should be compared. 

Section~\ref{SecCI} finally goes back to the incomplete information problem. We show that the hierarchical aggregation procedure discussed here can lead to a completely specified model (full knowledge of the joint distribution) whenever some additional conditions (conditional independence statements) between the risks are added. 

\section{Copula-Based Hierarchical Aggregation  of correlated risks}
\label{SectionAggregationTree}
In this section we describe the \textit{minimal} mechanism that enables one to formally compute the distribution function of the total portfolio in such a way that it fits with the idea of hierarchical aggregation described shortly in introduction. It is minimal in the sense that no additional hypotheses are needed to compute the total portfolio. In particular, \textit{the method below does not require the knowledge of the full joint distribution of the individual risks, nor does it specify it.}

The last section of the paper (Section~\ref{SecCI}) will discuss how extra conditions can be added to this minimal construction in order to define completely the joint distribution. As far as the diversification benefit is concerned, however, the distribution of the total portfolio is enough information and we do not need any additional knowledge on the joint distribution.

\subsection{Top down point of view: tree structure and summation decomposition}
Let $X_{i}$ be the individual risks (hereafter also called the leaves of the tree), and let $Z = \sum X_i$ be the total portfolio. We also denote with a bold face symbol the vector formed by these individual risks: $\mathbf{X}=(X_1,\ldots, X_n)$. 

The sum at risk associated to $Z$ is obtained by applying some risk measure on its distribution. Equivalently, and more conveniently for the present analysis,  the distribution of $Z$ is equally well given by its characteristic function:
\beq
\Phi_{Z}(t) \equiv \mathbb{E}\left[\exp\left(i t Z\right)\right]= \mathbb{E}\left[exp\left(i t \left(\sum X_i\right)\right)\right]
\eeq
There are different ways of computing it. Were the joint distribution of the $X_i$'s known, one could simply use the joint characteristic function $\Phi_{\mathbf{X}}(t_1,t_2,\ldots,t_n)= \mathbb{E}[\exp(i \mathbf{t}.\mathbf{X})]$ to get:
\beq
\Phi_{Z}(t)=\Phi_{\mathbf{X}}(t,t,\ldots,t) 
\eeq
This way of computing the sum corresponds to a direct aggregation of the risks, i.e. to a \textit{one-level} tree (hereafter also called a flat tree): 
\begin{figure}[ht]
\begin{center}
\includegraphics[width=10cm]{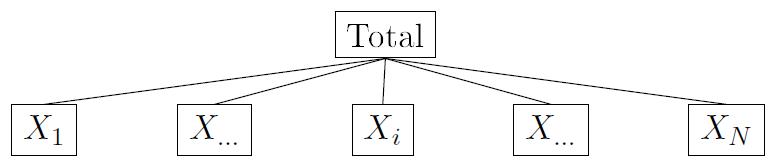}
\end{center}
\end{figure} 

Imagine however the case where one does not have all the information on the joint distribution function, but only partial information on, e.g. some joint distributions between \textit{subsets} of the leaves. For instance, split the vector of the $X_i$'s in two pieces $(X_1,\ldots,X_j)$ and $(X_{j+1},\ldots,X_N)$, and define $Y_1= X_1+ \ldots + X_j$ and $Y_2=X_{j+1} + \ldots +X_N$. Then, as a matter of fact, the characteristic function of $Z$ can also be written as:
\begin{equation}
\Phi_{Z}(t) = \E\left[\exp\left(i t \left(\sum X_i\right)\right)\right]=\mathbb{E}\left[\exp\left( (i \left(t Y_1 + t Y_2\right)\right)\right]\equiv\Phi_{Y_1,Y_2}(t,t)
\end{equation}
where we introduced the characteristic function $\Phi_{Y_1,Y_2}(t_1,t_2)$ that describes the joint distribution of the variables $Y_1$ and $Y_2$, which are partial sums of the leaves. This shows that the distribution of $Z$ might be computed even if the joint distribution of $\mathbf{X}$ is not known. Here it would be enough to know the joint distribution of $Y_1$ and $Y_2$. Of course we could have split the sum $Z$ in more than two terms. This way of computing $Z$, via a decomposition of the sum into pieces for which one does have some information exactly corresponds to the (intuitive) idea of aggregating risks in an hierarchical tree\footnote{Except that the point of view is top-down, whereas aggregation intuitively is bottom-up, see next subsection.}. In the above example, the tree is rather simple with two levels:
\begin{figure}[ht]
\begin{center}
\includegraphics[width=10cm]{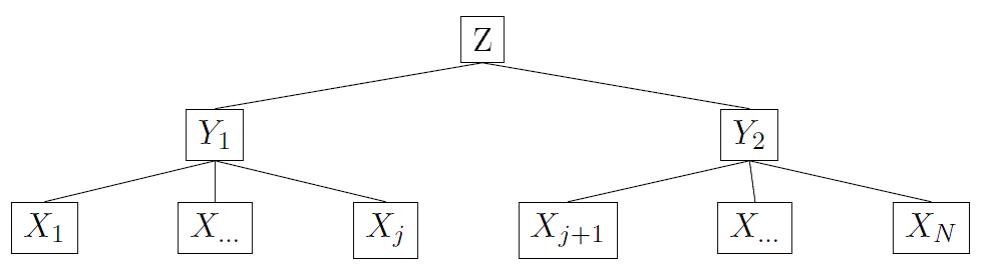}
\end{center}
\end{figure} 

Of course the procedure could iterate to fit more complex \textit{aggregation trees}. For example the joint distribution of the $(X_1,\ldots,X_j)$ might be unknown (and so would be $Y_1$), and one should therefore split further this vector, as, e.g. $Y_1=W_1+W_2$ with $W_1=(X_1+ \ldots + X_h)$ and $W_2=(X_{h+1}+ \ldots + X_j)$. Then $Y_1$ is calculable, provided the joint distribution of $W_1$ and $W_2$ is known.

\subsection{Bottom-up point of view: tree structure and risk aggregation}
As discussed in the introduction, in risk management analysis we are more concerned with a ''bottom-up'' point of view, where the marginals of the individual risks are usually known (from historical data, modelling, or any other methods), but their joint distribution is poorly known. Copulas, as we argued in Introduction, are well-known and powerful tools that enable to define a joint distribution function from given marginals \cite{Nelsen,EMBR}. Copulas originate from Sklar's theorem \cite{Sklar}, which essentially states that for any continuous distribution function $F$ with marginals $F_1, \ldots, F_d$, there exists one and only one function $C: [0,1]^d \rightarrow [0,1]$ such that $F$ can be written as
\beq
\label{Sklar}
F(x_1,\ldots,x_d)=C(F_1(x_1),\ldots,F_d(x_d))
\eeq
One is therefore able to construct dependency structures (joint distributions) starting from some given marginals $F_i$ and using appropriate functions $C$ (there are conditions on $C$ for it being a copula, see e.g. \cite{Nelsen}).

In the bottom-up point of view, one first split the leaves $X_i$ in $r$ disjoint sets, on the basis that (e.g.) those risks belonging to the same set share some common features between each other, and therefore are correlated. Then one joins together the leaves in these sets via some multidimensional copulas $C_s$ for $s \in [1,r]$. This, according to Sklar's theorem, defines their joint distributions. One thus ends up with $r$ multivariate distributions, and the partial sum of these risks $Y_1, \ldots, Y_r$ can be computed directly. Next level of aggregation consists in identifying again disjoint sets in the set $Y_1, \ldots, Y_r$, and proceeding similarly thereafter. Aggregation stops when the last copula ties together all the partial sum. 

It should be clear to the reader that the above bottom-up procedure fits perfectly well with the previous discussion (top-down approach), because once the copula is given, so is the joint distribution, and so is also the joint characteristic function. Let us however emphasize again that the procedure does not assume the knowledge of (nor specify) the full joint distribution. As explained, this does not spoil our ability to compute the distribution of the total portfolio.

\subsection{Regular Hierarchical Trees: Definition and notations}
In the following we will refer to trees in the usual sense of graph theory. Moreover we will restrict ourselves, for simplicity, to ``regular'' trees. 
\begin{definition}[Regular Hierarchical Trees]
\label{HTDef}
Let $k\geq 2$ and $m\geq 1$. A tree is said to be a $(k,m)$ regular tree if each node of the tree except the leaves have a common number $k \geq 2$ of children. Each layer (or level) of the tree is labeled by an integer $p$ ranging from $0$ to $m$. By definition the root corresponds to the level $0$ while the leaves populate the level $m$.
\end{definition}
The number $N_{(p)}$ of nodes at level $p$ is therefore $N_{(p)}=k^p$, and the total number of leaves, in particular is  $N_{(m)}=k^m$. In all the following, super or lower script in parenthesis $(m)$ or $(p)$ will always label the layer of the tree. 

In our construction, the tree is meant to be a useful graphical representation of random variables and their relationship. Therefore we also need to assign random variables to the nodes of the tree. We adopt the following conventions: The leaves of the tree are the individual risks. They are represented by random variables denoted by $X_i^{(m)}$, for $i \in [1,k^m]$. Similarly, at each level $p$, the nodes of the tree represent random variables denoted by $X_i^{(p)}$. 

Because of the aggregation mechanism, all these random variables are not independent from each other. At the contrary, any random variable associated to a node is given by the sum of the random variables associated to its child (except for the leaves). Moreover, hierarchical aggregation, as discussed in the introduction, also means assuming some dependencies between the nodes via the use of multivariate copulas (here $k-$dimensional copulas). Hence we have the following definition:

\begin{definition}[Hierarchical Aggregation Mechanism]
\label{HADef}
Let $\mathcal{T}$ be a $(k,m)$ regular tree as defined above. Let $p \neq m$, and let $\mathcal{J}_i^{(p)}$ be the set of child of the node $X_i^{(p)}$. Then, the aggregation mechanism refers to the following.  For all $p \neq m$ and relevant indices $i$, the $k$ nodes which belong to $\mathcal{J}_i^{(p)}$ 
\begin{enumerate}
\item Are joined together via some $k-$dimensional copula $C$ according to Eq.~(\ref{Sklar})
\item Are summed up to form their parent:
\beq
X_i^{(p)}=\displaystyle\sum_{s \in \mathcal{J}_i^{(p)}} X_s^{(p+1)}
\eeq 
\end{enumerate}
\end{definition}

Applying iteratively the above formula enables one to cascade down the full tree so that, in particular, the root of the tree corresponds to the full portfolio $Z\equiv X_1^{(0)}=\sum_i X_i^{(m)}$.

\section{Diversification benefit and risk measure: definitions}
\label{SecDB}
This section provides precise definitions for the diversification benefit. As discussed briefly in the Introduction, this requires a risk measure, see e.g.\cite{EMBR}. We advocated briefly the use of the Expected Shortfall or Tail-Value-At-Risk at some threshold $\alpha$. Actually we will use a slightly different risk measure. We will consider the $\textrm{xTVaR}$, which is difference between the $\textrm{TVaR}$ and the expectation value, because it is more representative of the value-at-risk for a company. 
Thus, $\textrm{xTVaR}_{\alpha} \equiv \E[X]-\textrm{TVaR}_{\alpha}[X]$ for a random variable $X$, whereas, as usual, the $\textrm{TVaR}$ is defined by
\beq
\textrm{TVaR}_{\alpha}[X] \equiv \frac{1}{\alpha} \int_{0}^{\alpha} \textrm{VaR}_{u}[X] du
\eeq
when it exists, and where $\textrm{VaR}_{u}$ is the Value-At-Risk at level $u$. Typically we choose $\alpha=0.01$, although the results for the Gaussian Tree will not depend on $\alpha$, see Section~\ref{SecGauss}.

We then define three sums at risk, two of which correspond to fictive extreme cases (zero or full dependency between the risks):
\begin{enumerate}
\item The sum at risk ``standalone'', which is the sum of all the susm at risk for each individual risks $X_i$: $S^{1}_{Z}=\sum_i \text{xTVaR}_{\alpha}(X_i)$. This is also the sum at risk of the full portfolio when all the individual risks are fully dependent between each other.
\item The actual sum at risk: $S_{Z}= \text{xTVaR}_{\alpha}(Z)= \text{xTVaR}_{\alpha}(\sum_i X_i)$. This quantity depends on the dependencies that exist between the individual risks.
\item The sum at risk while assuming that the individual risks $X_i$ are independent from each other:  $S_{Z}^0= \text{xTVaR}_{\alpha}(Z) |_\textrm{no dependency}$
\end{enumerate}
Clearly the actual sum at risk $S_{Z}$ can range from $S_{Z}^0$ (no dependencies, maximal diversification, minimal sum at risk) to $S^{1}_{Z}$ (full dependency between the individual risks, zero diversification and maximal sum at risk). Therefore we have $S_{Z}^0 \leq S_{Z} \leq S^{1}_{Z}$ and this suggests to define a dimensionless coefficient $\eta$, as 
\beq
\label{DefEta}
\eta = \frac{S_{Z} - S_{Z}^0}{S^{1}_{Z} - S_{Z}^0}
\eeq
This parameter, that we call the \textit{diversification factor} ranges from 0 to 1 and measures how close is the actual portfolio to the zero dependency case (high diversification  and $\eta \sim 0$)  or to the full dependency case (low diversification  and $\eta \sim1)$. Yet another instructive quantity is the diversification benefit itself, $DB$, which is usually defined as:
\beq
\label{DefDB}
\textrm{DB}= 1- \frac{S_{Z}}{S_Z^{1}}
\eeq

\section{Diversification in Hierarchical Gaussian Trees}
\label{SecGauss}
In this section we study and solve completely the hierarchical regular Gaussian tree, compute its diversification, and study its behavior with respect to the shape of the tree. Section~\ref{GaussSetup} precises the setup, Section~\ref{Main} gives the total portfolio as a result of the aggregation, Section~\ref{GaussDB} gives the diversification of the Gaussian Tree, and Section~\ref{GaussShape} discusses the influence of the shape of the tree on diversification.

\subsection{Definition of the Gaussian Tree}
\label{GaussSetup}
We study a regular $(k,m)$ aggregation tree as defined in definitions \ref{HTDef} and \ref{HADef}. We further simplify the problem by assuming the following:
\begin{assumption}[Individual risks are Gaussian]
The individual risks at the bottom of the tree are all modeled with the same univariate Gaussian with zero mean\footnote{The means are not relevant since we use the xTVaR risk measure and can therefore be set to $0$ without loss of generality.} and variance $\sigma_{(m)}^2$: 
\beq
\forall i, \quad X_i^{(m)} \sim \mathcal{N}(0,\sigma_{(m)}^2)  
\eeq 
\end{assumption}
\begin{assumption}[Gaussian copula is used]
 Any aggregation step  (see definition  \ref{HADef}) consists in tying together $k$ nodes via the $k-$dimensional equicorrelation Gaussian copula of dependency parameter $\rho$. This means that the $k \times k$ correlation matrix defining the Gaussian copula reads:
\beq
\label{Equi}
 \Sigma=  \mathbb{1}_k + \rho (J_k -\mathbb{1}_k),
\eeq
where $\mathbb{1}_k$ refers to the identity matrix of rank $k$, $J_k$ is the $k\times k$ matrix everywhere filled by 1, and the condition $-\frac{1}{k-1} < \rho < 1$ is assumed to hold for definiteness. 
\end{assumption}

Appendix \ref{GaussianWorld} provides some useful and well-known material for handling multivariate normals and gaussian copulas. 

\subsection{Result of the aggregation process}
\label{Main}
 The aggregation process in the gaussian tree results in the following:
\begin{theorem}
\label{Maintheo}
Any node of the regular gaussian tree is an univariate normal $X_i^{(p)} \sim \mathcal{N}(0,\sigma_{(p)}^2)$ with zero mean and standard deviation $\sigma_{(p)}$ given by:
\beq
\label{mainFor1}
\sigma_{(p)}= \sigma_{(m)} \left(k+(k^2 -k) \rho \right)^{(m-p)/2}
\eeq
In particular the total portfolio $Z$ is  $Z \sim \mathcal{N}(0,\sigma_Z^2)$ with
\beq
\label{mainFor}
\sigma_Z= \sigma_{(m)} \left(k+(k^2 -k) \rho \right)^{m/2}
\eeq
\end{theorem}
\textit{Proof:}
Let us start with the first aggregation step. Let $N=k^m$. At the bottom of the tree, there are $N/k$ groups of $k$ leaves that are tied together via the equicorrelation gaussian copula. Consider for instance the first group $X_1^{(m)}, \ldots, X_k^{(m)}$. By definition of Gaussian copulas, these $k$ univariate normals which are tied together via a Gaussian copula with correlation matrix $\Sigma$ have a joint distribution which is a $k-$variate normal with covariance matrix $\sigma_{(m)}^2 \times \Sigma$ (see Appendix \ref{GaussianWorld}). Then it follows from corollary \ref{corosum} that the sum $X_1^{(m-1)} \equiv X_1^{(m)}+\ldots+X_k^{(m)}$ is an univariate normal with variance given by Eq.~(\ref{SigmaAggreg}) and a mean given by the sum of the means (hence zero here). Applying the result to the equicorrelation matrix defined in Eq.~(\ref{Equi}) shows that  the sum $X_1^{(m-1)}$  is given by 
\beq
X_1^{(m-1)}=X_1^{(m)}+\ldots+X_k^{(m)} \sim \mathcal{N}(0, \sigma_{(m-1)}^2) 
\eeq
with 
\beq
\label{inter}
\sigma_{(m-1)}=\sigma_{(m)} \left(k+\left( k^2-k\right)\rho\right)^{1/2}
\eeq
This reasoning holds for all the other groups of $k$ nodes at the bottom of the tree. The level $(m-1)$ of the tree is thus completely determined: it is populated by the $X_i^{(m-1)}$ for $i \in k^{m-1}$, which are all univariate normal with zero mean and variance given by Eq.~(\ref{inter}). By definition of the aggregation mechanism, the process then repeats to upper levels. The theorem thus follows from a trivial iteration, and by solving the recurrence equation for the variances Eq.~(\ref{inter}) $\Box$

\subsection{Diversification in Gaussian Trees}
\label{GaussDB}
Based on theorem \ref{Maintheo} we compute the three sum at risks $S_0$, $S_Z$ and $S_1$ as defined in Section~\ref{SecDB} and then compute $\eta$ and $\textrm{DB}$. The full portfolio is univariate normal with zero mean and standard deviation $\sigma_Z$. A well-known result states that the $\textrm{TVaR}_{\alpha}$ of univariate normals $\mathcal{N}( \mu,\sigma^2)$ with mean $\mu$ reads $\textrm{TVaR}_{\alpha} = \mu + f(\alpha) \sigma$ for some function $f$, see e.g. \cite{EMBR}. As a consequence the $\textrm{xTVaR}$ simply reads $ f(\alpha) \sigma$.

It then follows that the sum at risk $S_Z$ reads $ f(\alpha) \sigma_Z$. Taking the limit for $\rho \to 0$ and $\rho \to 1$, we finally get:
\begin{eqnarray}
S^0_Z&=& f(\alpha) \sigma_{(m)} k^{m/2} \nonumber\\
S^Z&=&  f(\alpha) \sigma_{(m)} \left(k+\left( k^2-k\right)\rho\right)^{m/2} \nonumber\\
S^1_Z&=&  f(\alpha) \sigma_{(m)} k^m \nonumber
\end{eqnarray} 
One easily derives the diversification benefit
\beq
\label{DBGauss}
\textrm{DB}(k,m,\rho)=1- \left(\frac{1}{k}+\left(1 - \frac{1}{k}\right) \rho \right)^{m/2}
\eeq
and the diversification factor 
\beq
\eta(k,m,\rho)=\frac{\left(k+(k^2 -k) \rho \right)^{m/2} -k^{m/2}}{k^{m}-k^{m/2}}
\eeq
as functions of $k$ (width of the tree), $m$ (depth of the tree), and $\rho$ the dependency. We also recall that the number of individual risks is given by $N=k^m$. Notice that $f(\alpha)$ has dropped from the final result.

\subsection{``Thin'' trees diversify better than ``fat'' trees}
\label{GaussShape}

We now discuss the behaviour of the diversification as a function of the shape of the tree. Next figure Fig.(\ref{etavarm}) shows a family of curves that represent the behaviour of $\eta$ as a function $\rho$, for different values of $m$ but a fixed $k=3$.
\begin{figure}[ht]
\begin{center}
\includegraphics[width=10cm]{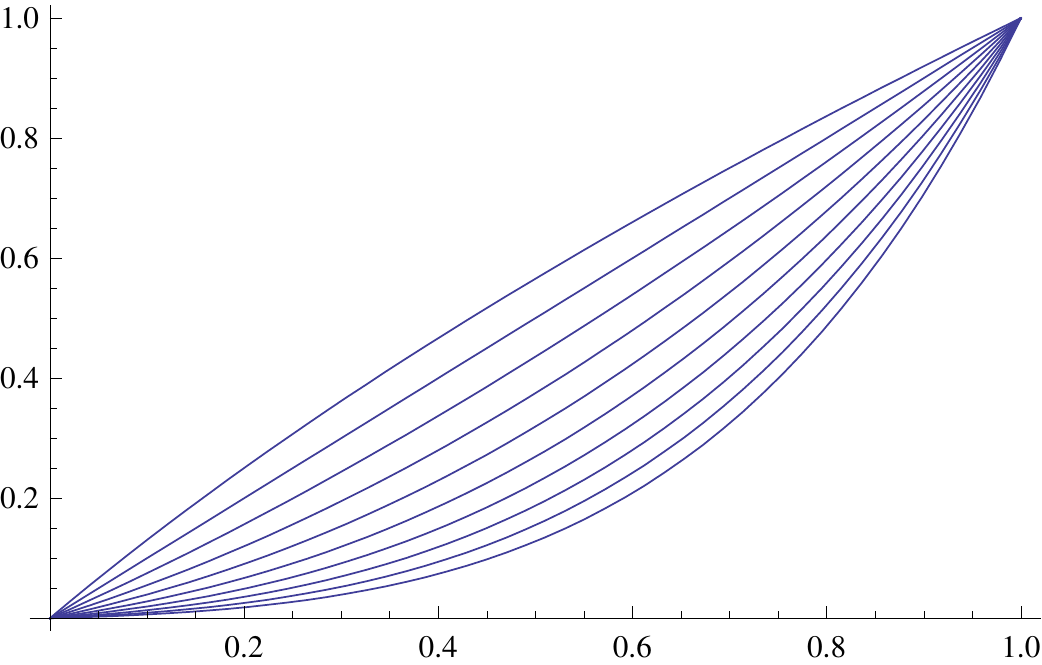}
\end{center}
\caption{Typical behaviour of $\eta$ (y-axis) with $\rho$ (x-axis) in a $(k,m)$ Gaussian tree. Here $k$ is set to $3$ and $m$ ranges from $1$  (top curve) to $10$ (bottom curve). The $m=1$ case corresponds to the flat tree with a concave curve for $\eta$. However, the curves become convex for $m \geq 3$, hence giving a small $\eta$ even if $\rho$ is large. 
}
\label{etavarm}
\end{figure} 
Notice that the number of leaves is not constant from one curve to the next. Observe in  Fig.(\ref{etavarm}) how increasing the depth of the tree makes the curves for $\eta$ more and more convex, whereas the curve for the flat tree $m=1$ is concave. 

This means that, for a given $\rho$, \textit{the hierarchical tree diversifies much better than the flat tree} (since $\eta$ is smaller). Moreover convexity implies that in hierarchical trees \textit{the diversification remains high even for large values of $\rho$}.

One may object that this simply comes as a consequence of increasing number of individual risks. However, we now show that the shape of the tree does also play a role \textit{per se} in the diversification of risks. Whenever $N$ is fixed, $m$ and $k$ are related through $N=k^m$. Qualitatively speaking, therefore, the difference between two trees $(k,m) \neq (k',m')$  is a sort of opening angle, and we have the following result:

\begin{proposition}[Thin trees diversify better than fat trees]
Let $(k,m)$ and $(k',m')$ such that  $k'^{m'}=k^m=N$. Then $\textrm{DB}'< \textrm{DB}$ if $k'>k$.
\end{proposition}
Thus the diversification benefit is smaller for trees with larger width $k$. The proof is straightforward. One writes the diversification benefits as $\textrm{DB}'=1-q'$ and $\textrm{DB}=1-q$ and forms the ratio
\beq
\ln(q'/q)=\frac{m}{2} \left( \ln\left(\frac{1}{k}(1-\rho)+\rho\right)- \frac{\ln k}{\ln k'}\ln\left(\frac{1}{k'}(1-\rho)+\rho\right)\right)
\eeq
which shows that $q'>q$ when $k'>k$ $\Box$

The following example illustrates this. We take e.g. $\rho=0.4$, and $k=2$, $m=10$, $N=1024$. Then we find $DB \approx 83\%$ and $\eta \approx 0.14$. However, with $k=4$, $m=5$, $N=1024$ we find $DB \approx 77\%$ and $\eta \approx 0.2$. 

To make some contact with the numerical studies done in Section~\ref{SecLN}, next Figure \ref{dbgauss} also compares the diversification benefits for the flat Gaussian tree ($k=729,m=1$) and for some hierarchical Gaussian tree ($k=3,m=6$). As Figure \ref{dbgauss} shows, the hierarchical tree diversifies much more than the flat tree.
\begin{figure}[ht]
\begin{center}
\includegraphics[width=10cm]{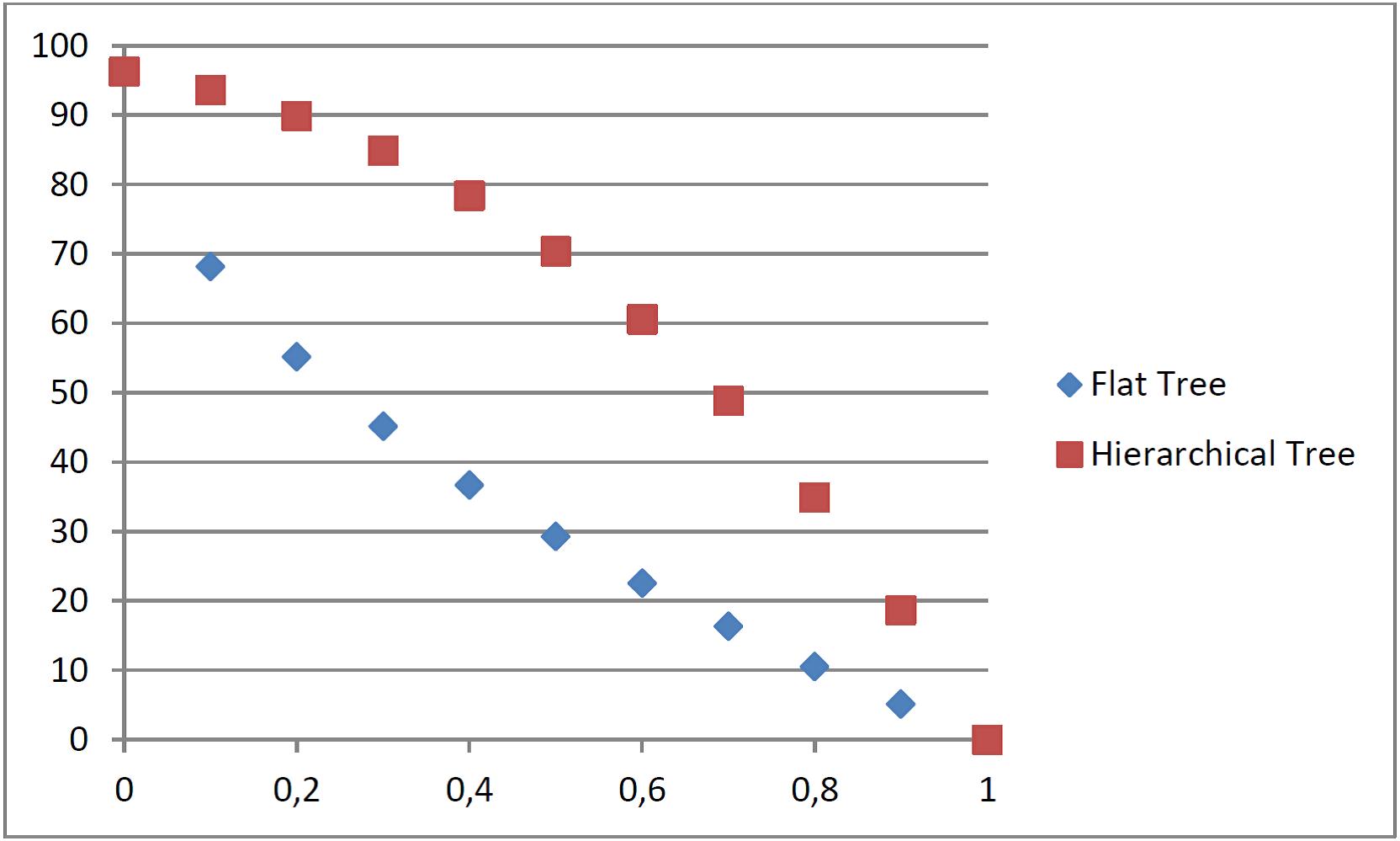}
\end{center}
\caption{Diversification benefit in percents as a function of $\rho$ for both the flat and hierarchical Gaussian trees with parameter as given in the text.}
\label{dbgauss}
\end{figure} 

This result regarding the influence of the shape of the tree on its diversification shall not come as a surprise. As discussed briefly in the introduction indeed, it is quite intuitive because individual risks are typically dependent on each other via some upper node in the tree, i.e. after several times the application of the Gaussian copula. This has the tendency to lower the effective correlation between these risks, and this happens more often in thin trees than in fat trees. Equation  (\ref{DBGauss}) illustrates this with the emergence of powers of $\rho$ in the result for diversification. 

The above discussion cannot be more than qualitative. Indeed the joint distribution of the leaves is not completely known and we cannot compute the correlation coefficients between distant individual risks. Section~\ref{SecCI} will however show that when one adds some conditions that enable to specify fully the distribution, then one is able to compute these effective couplings, and show that they decrease with the distance between the individual risks in the tree.

\section{Diversification in Hierarchical LogNormal Treess}
\label{SecLN}
In this section we add to the full treatment given in the Gaussian case some numerical results on LogNormal trees. By this we mean hierarchical regular trees as defined previously, but the individual risks are now described by the some common LogNormal distribution. Our aim is to see whether the previous conclusions in the Gaussian case also hold in more realistic\footnote{LogNormal distributions are widely used in risk modelling in insurance companies, see \cite{BurgiDacorogna}} settings.

\subsection{Methodology and parameterization of the tree}
 In this section we set $k=3$ and $m=6$, i.e. we consider a tree which aggregates together $N=3^6=729$ individual risks. The aggregation has been done by means of a Gaussian copula of parameter $\rho$, and also with a Clayton copula of parameter $\theta$. Diversification benefit $\textrm{DB}$ and diversification factor $\eta$ have been evaluated numerically with a threshold $\alpha=0.01$, using the IGLOO software \cite{Igloo}. Results are systematically compared to the diversification of a flat tree $k=729,m=1$.

The results show that the hierarchical tree diversifies much better than the flat tree. Moreover results show that the lognormal hierarchical tree also has the tendency to make more convex the curve for the diversification factor $\eta$, especially for Gaussian copulas, see graphs below. This establishes the relevance of the Gaussian tree as a toy-model of aggregation trees.

The parameters of the LogNormal have been chosen in order to fit realistic portfolios. Explicitly, by imposing a mean of around $670'000$ and a standard deviation of $8.1$ millions for these LogNormals, then one finds that the maximal sum at risk (full dependency) reads $S^1_Z=22.4$ (in bn), and the minimal sum at risk (zero dependency) is approximately $S^0_Z=1.3$ bn.

\subsection{LogNormal Tree: Gaussian copula}
The dependencies are set by a Gaussian copula with dependency parameter $\rho$. We computed numerically the sum at risk $S_{Z}$ as a function $\rho$, for both the $(k=3,m=6)$ hierarchical tree and the flat $(k=729,m=1)$ tree.

\begin{figure}[ht]
\begin{center}
\includegraphics[width=10cm]{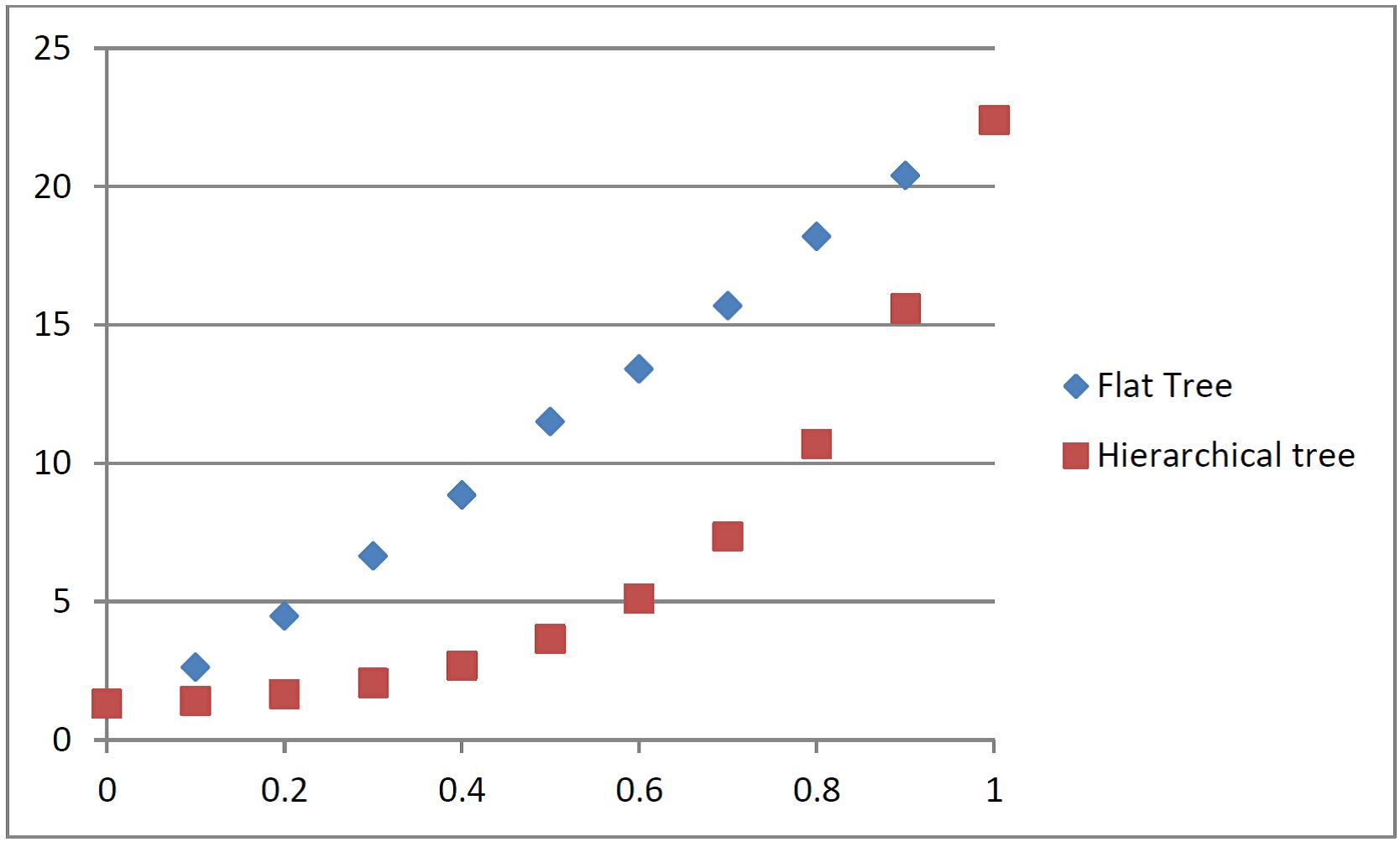}
\end{center}
\caption{LogNormal Tree with Gaussian copula. The xTVaR at $99\%$ of the full portfolio, in billions, as a function of $\rho$. This interpolates from $S^0_Z=1.3$ to $S^1_Z=22.4$ bn. }
\label{srho}
\end{figure}

\begin{figure}[ht]
\begin{center}
\includegraphics[width=10cm]{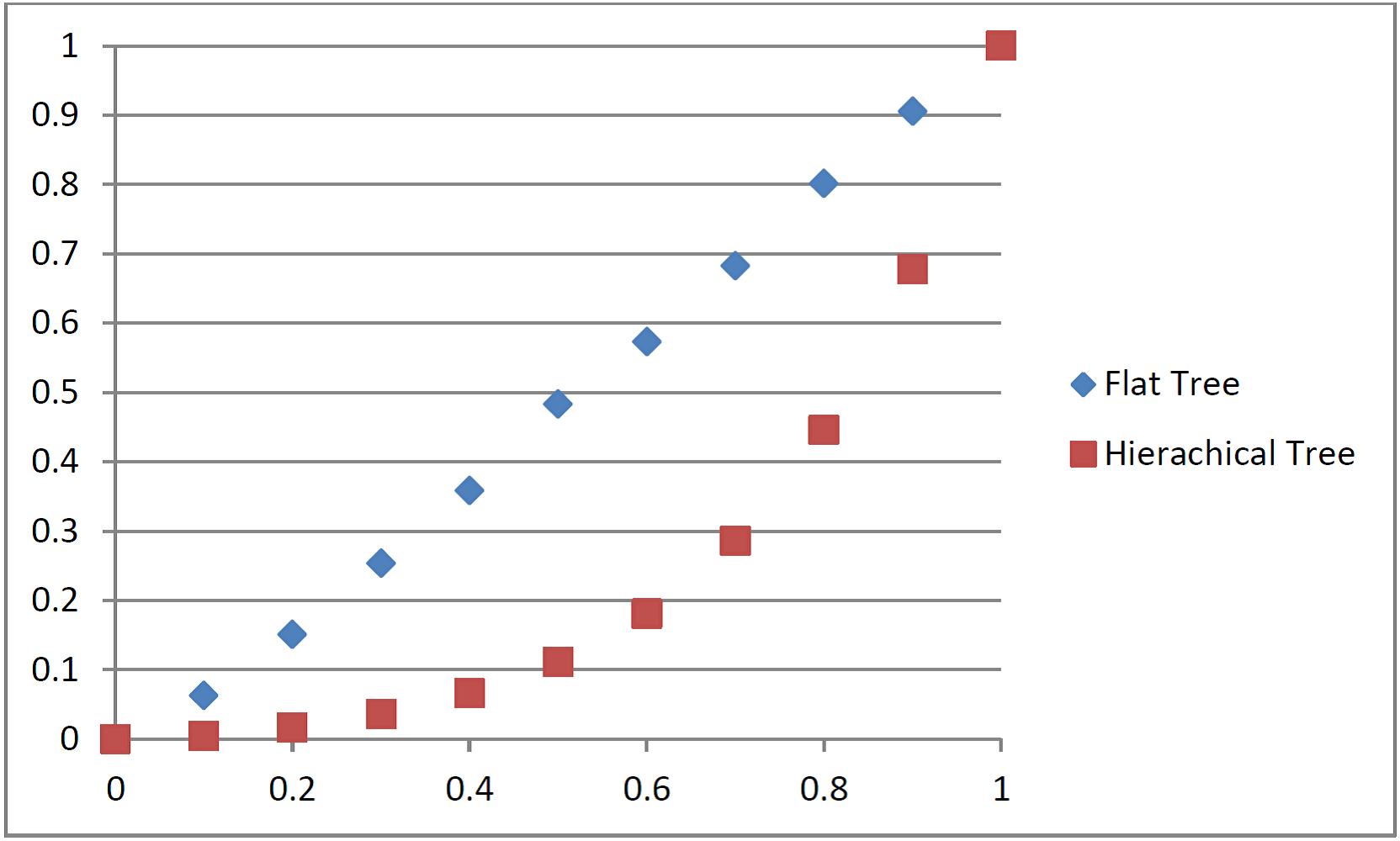}
\end{center}
\caption{LogNormal Tree with Gaussian copula. The factor $\eta$ as a function of $\rho$. It is essentially a rescaling of the previous curve. }
\label{etarho}
\end{figure}

\begin{figure}[ht]
\begin{center}
\includegraphics[width=10cm]{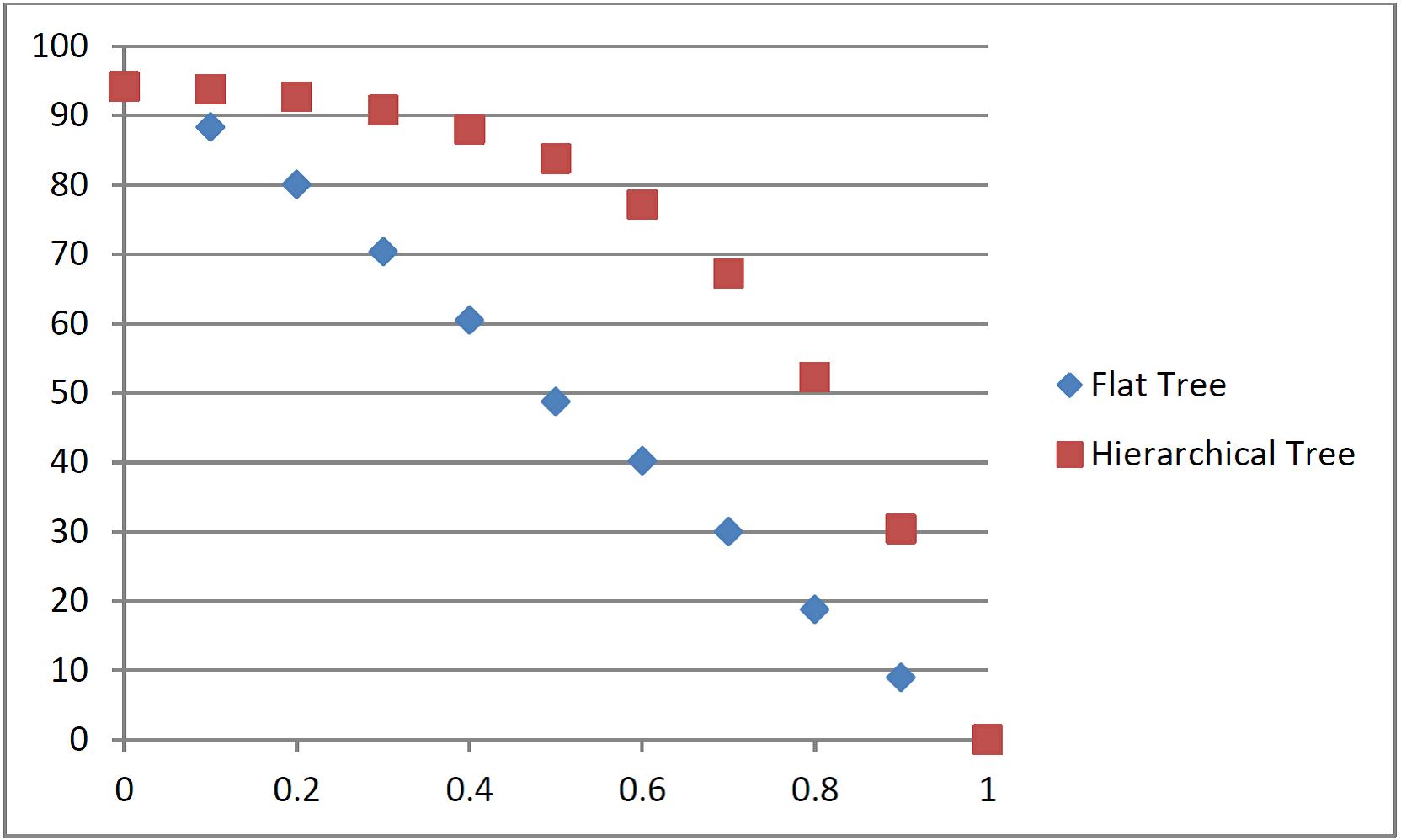}
\end{center}
\caption{LogNormal Tree with Gaussian copula. The diversification benefit in percents as a function of $\rho$.}
\label{dbrho}
\end{figure}

Fig. (\ref{srho}) shows the sum-at-risk as a function of the dependency parameter $\rho$. Similarly Fig. (\ref{etarho}) and
Fig. (\ref{dbrho}) respectively show the behaviour $\eta$ and $\textrm{DB}$. We see that, as in the Gaussian tree, the following holds in the lognormal tree with Gaussian copula:
\begin{itemize}
\item  The sum-at-risk stays almost constant for small values of $\rho$ in the hierarchical tree.
\item The curve for the $\eta$ becomes convex when adding new layers to the tree.
\item The diversification benefit stays relatively constant for small values of $\rho$.
\end{itemize}
These properties all show that the successive applications of the Gaussian copula in the hierarchical tree almost completely suppress the dependencies for not too large values of $\rho$.

\subsection{LogNormal Tree: Clayton copula}
\begin{figure}[ht]
\begin{center}
\includegraphics[width=10cm]{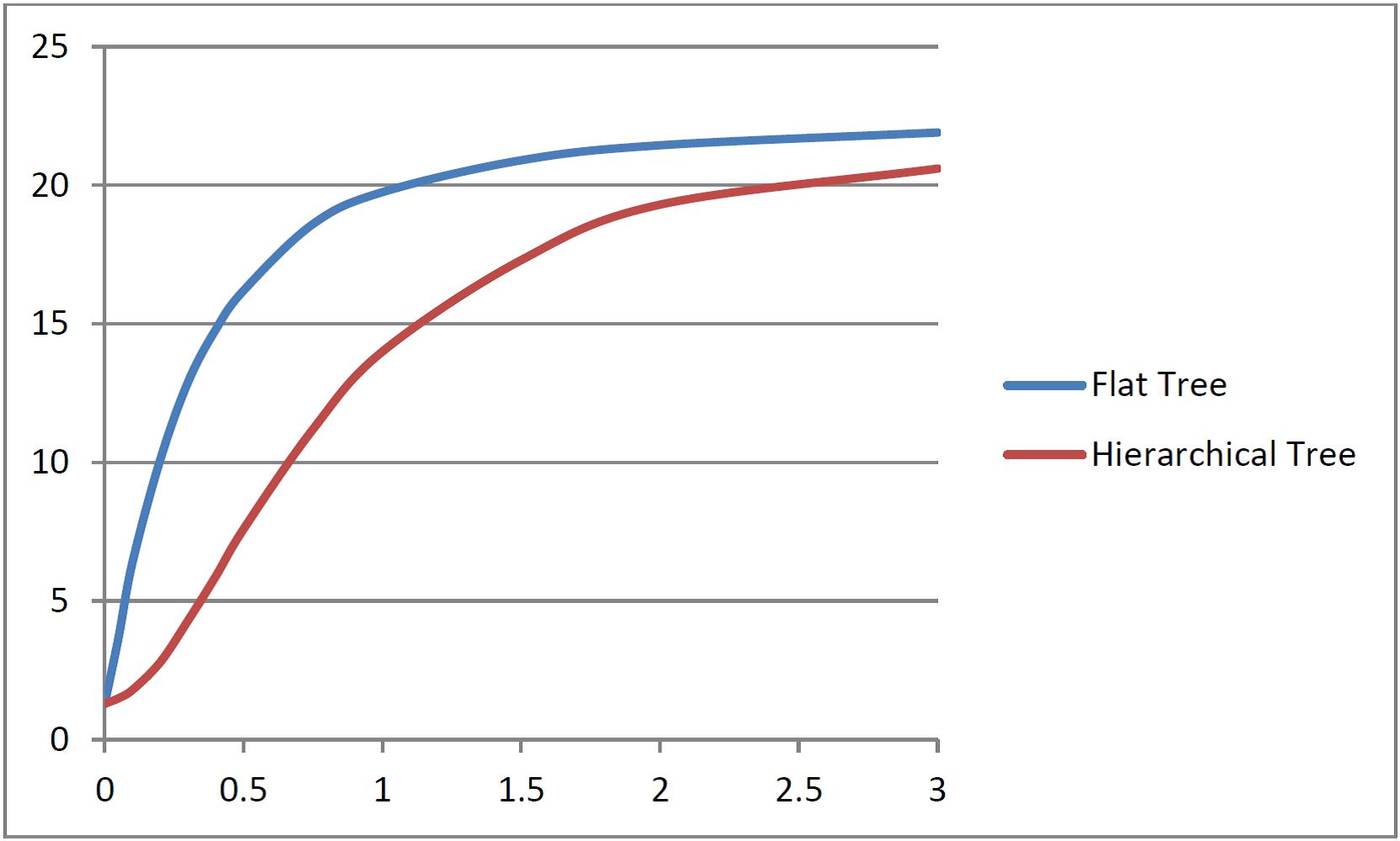}
\end{center}
\caption{LogNormal Tree with Clayton copula. The xTVaR at $99\%$ of the full portfolio, in billions, as a function of Clayton parameter.}
\label{sclayt}
\end{figure}

\begin{figure}[ht]
\begin{center}
\includegraphics[width=10cm]{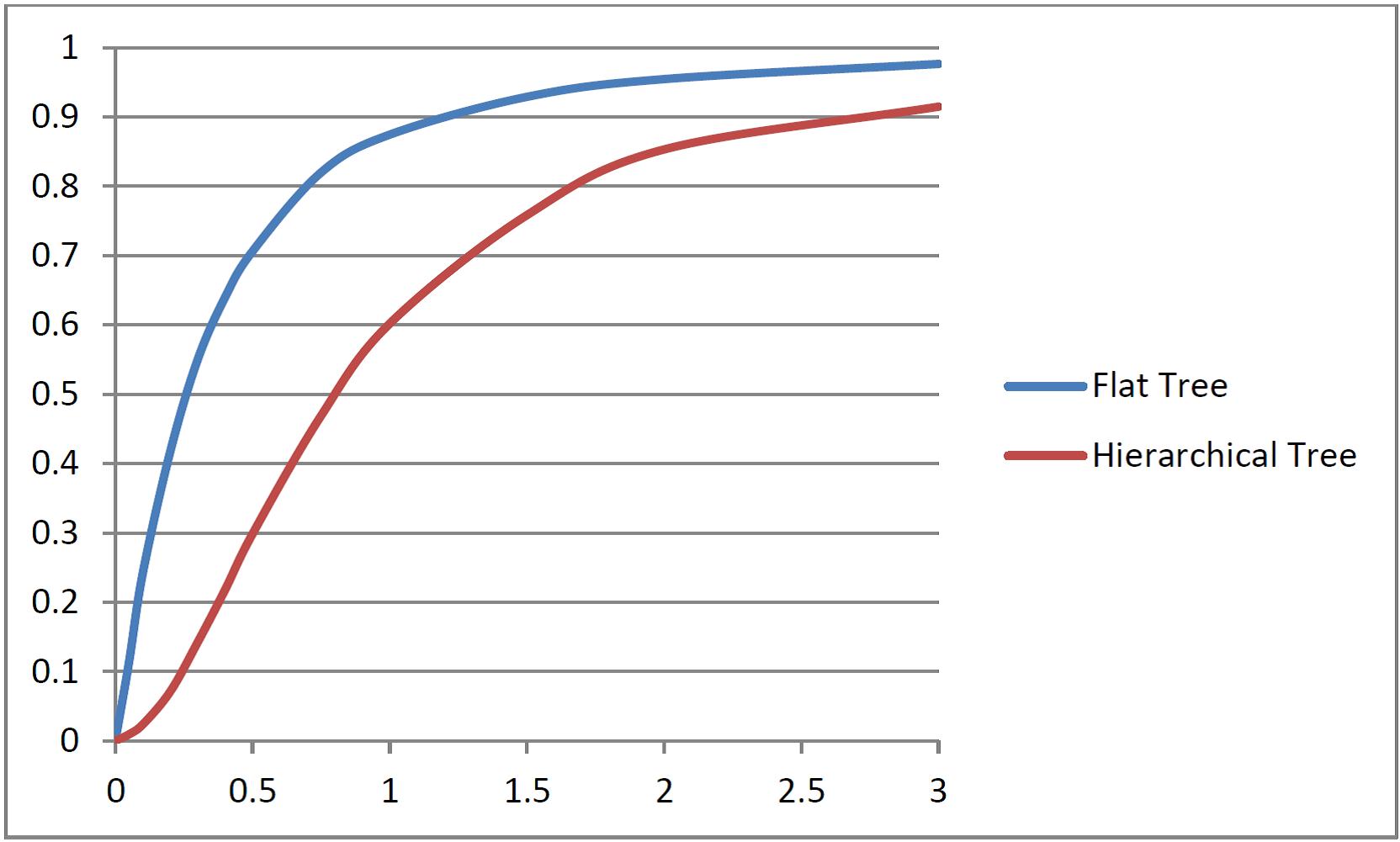}
\end{center}
\caption{LogNormal Tree with Clayton copula. The factor $\eta$ as a function of Clayton parameter.}
\label{etaclayt}
\end{figure}

\begin{figure}[ht]
\begin{center}
\includegraphics[width=10cm]{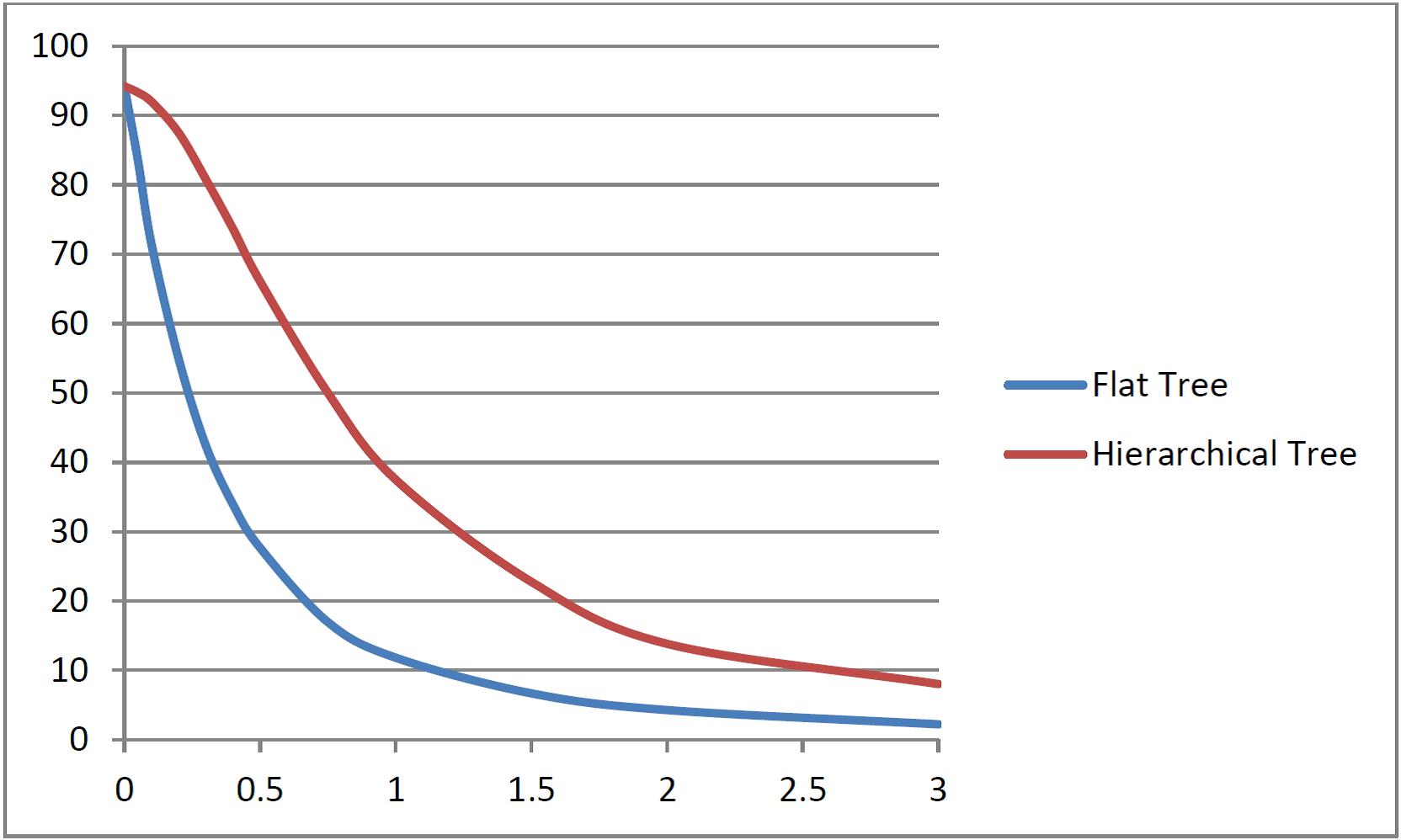}
\end{center}
\caption{LogNormal Tree with Clayton copula. The diversification benefit in percents, as a function of Clayton parameter.}
\label{dbclayt}
\end{figure}

We performed the same analysis but using this time a Clayton copula of parameter $\theta$, and varying its strength. Figures (\ref{sclayt}, \ref{etaclayt} and \ref{dbclayt}) respectively show (smoothed curves of) the sum-at-risk, $\eta$ and $\textrm{DB}$ as a function of the Clayton parameter $\theta$. 

We note that the general shapes of the curves are  rather different from the ones found using a Gaussian copula. The curves for the flat tree are much more concave (for the sum at risk and $\eta$) and the hierarchical aggregation mechanism fails partially to turn it into a convex curve. Still, the hierarchical tree diversifies much better than the flat tree. In other words also, the diversification benefit fails to stay almost constant for some range, meaning that the hierarchical tree is not able to balance completely the high dependency set by the Clayton copula. 


\clearpage
\section{Specifying fully the tree}
\label{SecCI}
We saw in Section~\ref{SectionAggregationTree} that the hierarchical aggregation mechanism does not fully specify the joint distribution but rather provides minimum amount of information to determine the full portfolio.

\subsection{Conditional Independent Trees}
Inspired by \cite[Chp.8]{Bluebook}, we now add some conditional independence statements to the regular Gaussian tree and show that it enables to specify the joint distribution of the leaves of the tree. 

Let us note $X \leq Y$ whenever $X$ is a child of $Y$. This defines a partial order relationship in the tree. Then we define:
\begin{definition}[Conditional independent trees]
\label{CItree}
A tree $\mathcal{T}$ is said to be conditional independent iff for all $X, Y, S \in \mathcal{T}$ such that  $X \leq Y$ and  $S \nleqslant Y$, then $(X\independent S) | Y $.
\end{definition}
Note that $ (X\independent S) | Y \Leftrightarrow (S\independent X) | Y$. Then we have the following theorem.
\begin{theorem}
\label{TheoCI}
There exist one and only one regular $(k,m)$ conditional independent Gaussian tree $\mathcal{T}_{k,m}$. The $N=k^{m}$ leaves of the tree are jointly normal with covariance matrix $\mathcal{C}^{(m)}$ given by the following recurrence relations:
\[
 C^{(0)} = 
 \begin{pmatrix}
 \sigma_Z^2
 \end{pmatrix}
\]

\[
 C^{(1)} = \sigma_{(1)}^2
 \begin{pmatrix}
  1 & \rho & \cdots & \rho \\
  \rho & 1 & \ddots & \vdots\\
  \vdots  & \ddots  & \ddots &  \rho  \\
\rho &\ldots & \rho &1
 \end{pmatrix}
\]
And, $\forall m \geq 2$:
\[
 C^{(m)} = \sigma_{(m)}^2
 \begin{pmatrix}
  \frac{C^{(m-1)}}{\sigma_{(m-1)}^2} &   B_{(m-1)}' & \cdots &   B_{(m-1)}' \\
  B_{(m-1)} &   \frac{C^{(m-1)}}{\sigma_{(m-1)}^2} & \ddots & \vdots\\
  \vdots  & \ddots  & \ddots &    B_{(m-1)}'  \\
    B_{(m-1)} &\ldots &   B_{(m-1)} &  \frac{C^{(m-1)}}{\sigma_{(m-1)}^2}
 \end{pmatrix}
\]
where $\mathcal{C}^{(m)}$ is here written as a $k\times k$ block matrix, with blocks of size $k^{m-1}$, and where
\begin{itemize}
\item The recurrence relations between the variances is:
\beq
\sigma_{(p-1)}=\sigma_{(p)} \sqrt{k+(k^2 -k) \rho} \nonumber
\eeq

\item The matrix $B_{(m-1)}$ is given by $B_{(m-1)} = \beta_{m-1} J_{k^{m-1}}$ with
\beq
\beta_{m-1}= \rho \left( \frac{1}{k} +\left( 1-\frac{1}{k}\right) \rho\right)^{m-1} \nonumber
\eeq
\end{itemize}
\end{theorem}
\textit{Proof.} See Appendix \ref{AppProof}

\subsection{``Effective'' correlations}
Results on the diversification in the Gaussian tree in Section~\ref{SecGauss}  made clear that applying successive Gaussian copula of parameter $\rho$ in the aggregation mechanism implies smaller and smaller correlation coefficients between nodes that are not directly connected. This can now be given an explicit formulation within the context of the conditional independent Gaussian tree.
 
\begin{proposition} 
\label{proposeff}
Two risks $X_{i}^{(m)}$ and $X_{j}^{(m)}$ which are first connected to each other via a node located at level $p$ have between them the linear correlation coefficient $\tilde{\rho}^{(m,p)} \equiv\textrm{cov}(X_{i}^{(m)},X_{j}^{(m)}) $ with:
\beq
\tilde{\rho}^{(m,p)} = \rho \left( \frac{1}{k} + \left( 1 -\frac{1}{k}\right) \rho\right)^{m-p-1}
\eeq
\end{proposition}
\textit{Proof.} See step 3 of Appendix \ref{AppProof}. The smallest correlation coefficients in the covariance matrix are of course between the most distant leaves, i.e. between leaves that are connected to each other only via the root of tree ($p=0$). This explains why the overall diversification benefit can be large (or that the $\eta$ factor can be small) even if the dependency parameter $\rho$ is large (see also Fig.(\ref{etavarm})), whenever $m$ is large enough. 

\section{Conclusions}
We provided a self-contained introduction to hierarchical aggregation of correlated risks within trees of aggregation. We discussed the advantages of such a method, discussed its relevance for financial industry and in particular insurers and reinsurers. The mathematics underlying the definition of a tree as a useful graphical representation has been. 

Thanks to an completely solvable model for aggregation trees, namely the Gaussian tree, we showed how the shape of the tree, via its impact on the dependency structure, influences the resulting diversification benefit. In particular we showed that the depth of the tree, or more precisely its thinness, heavily influences its diversification. 

More precisely, one of the main conclusion is that hierarchical trees have the tendency to lower effectively the dependency set at each step of the aggregation. As a consequence the diversification benefit stays relatively large and constant for small dependency parameter, and starts decreasing only for high values of the dependency parameter.

This should be understood as a systematic effect that should be kept in mind while modelling risk aggregation: although an high dependency is applied at each aggregation steps, the diversification might stay relatively large depending on the shape of the tree. This represents a risk for modelling: inadequate shapes of aggregation trees might lead to a large over/under estimation of the diversification gain.

The only way to avoid these systematic effects again relies on an adequate modeling of risks. As was already the case when we considered the non-trivial role of the order in which the risks are aggregated, it is critical to stay as close as possible to the business. The shape of the tree shall not be arbitrary nor based on oversimplifications. 

Of course real portfolio analysis is plagued by the fact that due to their complexity, there are always modelling choices ambiguities. A good practice, although much time consuming, would thus be to design and calibrate several distinct aggregation trees corresponding to different modeling choices, in order to get a better picture of the landscape of diversification.

In the last part of this work, we also showed how the joint distribution of the risks can be entirely determined in the Gaussian case by adding other constraints to the aggregation process, namely conditional independence statements between nodes in the tree. Further work would be needed to understand whether the set of constraints that we used is the minimal set that enables to specify completely the joint distribution. Also, it would be worth studying whether this holds as well for more general (non-Gaussian) trees. 

\acknowledgments
The author was supported by SCOR Switzerland AG when  part of this work was done, and wishes to thank F.Uffer and M. Dacorogna (SCOR Switzerland AG) for fruitful discussions.
%

\appendix
\section{Multivariate Normals and Gaussian Copula}
\label{GaussianWorld}

\begin{definition}
A random variable $X$ is said to be univariate normal, or normal for short, if its density function reads:
\beq
\label{jointnormal}
f(x)= \frac{1}{(2 \pi \sigma^2)^{1/2}} \exp\left(-\frac{(x -\boldsymbol\mu)^2}{2 \sigma^2} \right)
\eeq
and we note $X \sim \mathcal{N}(\boldsymbol\mu, \sigma^2)$
\end{definition}

\begin{definition}
A random vector $\mathbf{X}=(X_1, \ldots, X_N)$ is said to have a non-singular multivariate normal distribution iff there exists a symmetric positive-definite matrix $\mathcal{C}$ and a vector $\boldsymbol\mu$ such that the joint density reads
\beq
\label{jointnormal}
f(\mathbf{x})= \frac{1}{(2 \pi)^{N/2} \sqrt{\det(\mathcal{C})}} \exp\left(-\frac{1}{2} \mathbf{(x -\boldsymbol\mu)}' \mathcal{C}^{-1} \mathbf{(x -\boldsymbol\mu)}\right)
\eeq
where $\mathbf{x}$ and $\boldsymbol\mu$ are vectors in $ \mathbb{R}^N$
\beq
\mathbf{x}=\begin{pmatrix}x_1 \\ \ldots \\ x_N \end{pmatrix} \nonumber
\quad \boldsymbol\mu=\begin{pmatrix} \mu_1 \\ \ldots \\ \mu_N \end{pmatrix}, 
\eeq
and where the prime indicates the transposition. 
\end{definition}
Observe the boldface notation that, as usual, denotes vectors (and sometimes matrices as well). The matrix $\mathcal{C} = \textrm{cov}(\mathbf{X})$ is known as the covariance matrix, i.e. a $N \times N$ matrix whose elements are $\mathcal{C}_{ij} = \text{cov}(X_i,X_j)$.

\begin{theorem}[Linear combinations of multivariate normal] See e.g. \cite[Chp.3]{EMBR} \\
\label{theolin}
Let $\mathbf{X}=(X_1, \ldots, X_N)$ be any multivariate normal distribution of dimension $N$, with covariance matrix $\text{cov}(\mathbf{X})$, and let  $A$ be a $k \times N$ real matrix $A$. Then 
\begin{equation}
\mathbf{Y}=A \mathbf{X}' \nonumber
\end{equation}  
is also multivariate normal of dimension $k$  with $k \times k$ covariant matrix given by:
\begin{equation}
\label{CovSumGauss}
\text{cov}(\mathbf{Y})=A \, \text{cov} (\mathbf{X}) A' 
\end{equation}
\end{theorem}

As a consequence we have the following:
\begin{corollary}[Sums of Gaussians that are jointly normal] See e.g. \cite[Chp.3]{EMBR}\\
\label{corosum}
Let $\mathbf{X}=(X_1, \ldots, X_N)$ be any multivariate normal distribution of dimension $N$, as defined above, with covariance matrix $\mathcal{C}_{ij}$ and mean $\boldsymbol\mu$. Then the sum of these Gaussian marginals is itself univariate normal. Defining $Z =\sum_{i=1}^{i=N} X_i $, we have 
\beq
\label{distribsum}
Z \sim \mathcal{N}(\mu_Z, \sigma_Z^2) 
\eeq
with
\beq
\label{SigmaAggreg}
\sigma_Z^2=\displaystyle\sum_{i=1}^{N} \displaystyle\sum_{j=1}^{N} \mathcal{C}_{ij} \qquad \text{and} \quad  \mu_Z= \displaystyle\sum_{i=1}^{N} \mu_i
\eeq
\end{corollary}

We also recall a standard result on conditioning on multivariate normals. 
\begin{theorem}[Conditioning on multivariate normals.] See e.g. \cite[Chp.3]{EMBR}\\
\label{theocondition}
Let $\mathbf{X}=(X_1, \ldots, X_N)$ be jointly normal with covariance matrix $\mathcal{C}$ and mean $\boldsymbol\mu$. Let $\boldsymbol\mu$ and $\mathcal{C}$ be partitioned as follow: $\boldsymbol\mu=(\boldsymbol\mu_1,\boldsymbol\mu_2)$ with respective size $q$ and $N-q$, and accordingly
\[
\mathcal{C} = 
 \begin{pmatrix}
  \mathcal{C}_{11} & \mathcal{C}_{12}  \\
   \mathcal{C}_{21} & \mathcal{C}_{22}  \\
 \end{pmatrix}
\]
Then provided $\mathcal{C}_{22}$ is non-negative $\mathbf{X_1}|\mathbf{X_2}=\mathbf{x_2}$ is multivariate normal with mean $\tilde{\boldsymbol\mu}=\boldsymbol\mu_1+\mathcal{C}_{12} \mathcal{C}_{22}^{-1} (\mathbf{x_2} -\boldsymbol\mu_2)$ and covariance matrix $\tilde{\mathcal{C}}=\mathcal{C}_{11} -\mathcal{C}_{12} \mathcal{C}_{22}^{-1} \mathcal{C}_{21}$.
\end{theorem}
As a consequence we have the following
\begin{corollary}
\label{CoroCondition}
Let $\mathbf{X}=(X_1, \ldots, X_N)$ be jointly normal  with covariance matrix $\mathcal{C}$ and mean $\boldsymbol\mu$. Let $Y=\displaystyle\sum X_i$. Then $\mathbf{X}|Y=y$ is multivariate normal with mean $\tilde{\boldsymbol\mu}=\boldsymbol\mu+\mathcal{C}^*_{12} \mathcal{C}_{22}^{* -1} (y -\sum \mu_i)$ and covariance matrix  $\tilde{\mathcal{C}}=\mathcal{C} -\mathcal{C}_{12}^* \mathcal{C}_{22}^{* -1} \mathcal{C}_{21}^*$, where $\mathcal{C}^*_{12}$ is a $N$ column vector of elements $\sum_j\mathcal{C}_{ij}$ and $\mathcal{C}^*_{22}=\sum_{ij}\mathcal{C}_{ij}$.
\end{corollary}
\textit{Proof.} First consider the vector $\chi=(X_1, \ldots, X_N,Y)$. Note that it is also given by  $\chi = \mathbf{M}.\mathbf{X}$ where $\mathbf{M}$ is a $(N+1) \times N$ matrix (the upper $N\times N$ matrix is the identity matrix, and the last line of $\mathbf{M}$ is full of $1$). By virtue of theorem \ref{theolin}, it implies that $\chi$ is a $N+1$ multivariate normal with mean given by $\boldsymbol\mu^*=\mathbf{M}\boldsymbol\mu$ and covariance matrix $\mathcal{C}^*=\mathbf{M} \mathcal{C} \mathbf{M}'$. One computes $\boldsymbol\mu^*=(\boldsymbol\mu, \sum_i \mu_i)'$, and 
\[
\mathcal{C}^* = 
 \begin{pmatrix}
  \mathcal{C} & \mathcal{C}^*_{12}  \\
   \mathcal{C}^*_{21} & \mathcal{C}^*_{22}  \\
 \end{pmatrix}
\]
Note that $\mathcal{C}^*_{22} \geq 0$ because $\mathcal{C}$ is non-negative. Therefore we can apply the previous theorem \ref{theocondition} to show that $\mathbf{X}|Y=y$ is $N-$dimensional normal with $\tilde{\boldsymbol\mu}=\boldsymbol\mu+\mathcal{C}^*_{12} \mathcal{C}_{22}^{* -1} (y -\sum \mu_i)$ and $\tilde{\mathcal{C}}=\mathcal{C} -\mathcal{C}_{12}^* \mathcal{C}_{22}^{* -1} \mathcal{C}_{21}^*$ $\Box$

\begin{definition}[Gaussian Copula]
Let $\Phi_\Sigma$ be the cumulative distribution function of a $k-$variate normal with zero mean and covariance matrix $\Sigma$, and let $\Phi$ be the cdf of a standard normal. Then the Gaussian copula with correlation matrix $\Sigma$ is the function:
\beq
C_{\Sigma}^{\textrm{G}}(\mathbf{u})= \Phi_{\Sigma}\left( \Phi^{-1}(u_1), \ldots,  \Phi^{-1}(u_k)\right)
\eeq
\end{definition}
In the case where $\Sigma$ reads  $\Sigma=  \mathbb{1}_k + \rho (J_k -\mathbb{1}_k)$, i.e.
\[
\Sigma= 
 \begin{pmatrix} 
1 & \rho & \cdots & \rho \\
  \rho & 1 & \ddots & \vdots\\
  \vdots  & \ddots  & \ddots &  \rho  \\
\rho &\ldots & \rho &1
 \end{pmatrix}
\]
we will refer to the equicorrelation Gaussian copula of parameter $\rho$. It is also an exchangeable copula, meaning that $C_{\Sigma}^{\textrm{G}}$ is invariant for any permutations of the $u_i$'s.

\section{Proof of Theorem \ref{TheoCI}}
\label{AppProof}

Given any $k \geq 2$, the proof is by strong recurrence on $m$. Initialization ($m=1$) is clear (see e.g. proof of theorem \ref{Maintheo}). The basic idea behind the proof by recurrence is to note that $T_{k,m}$ is obtained by aggregating together $k$ times the subtrees $T_{k,m-1}$ for which the theorem is assumed to hold. We have then: $T_{k,m}=$
\begin{figure}[ht]
\begin{center}
\includegraphics[width=10cm]{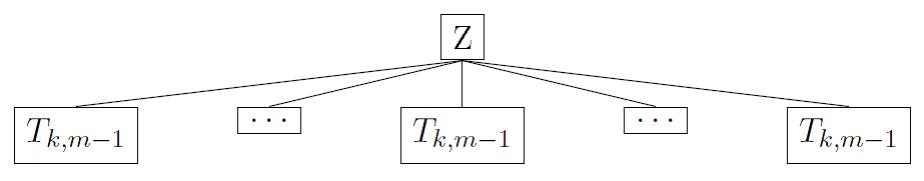}
\end{center}
\end{figure} 

 Note that in this approach it is necessary to rename accordingly all the random variables, variances and so on attached to the subtrees (e.g. $\sigma_{(m-1)}$ of the subtree becomes $\sigma_{(m)}$ of the full tree, etc). Also, for more clarity, we denote in this proof the roots of the subtrees by $Y$:
\beq
Y_i \equiv  X_{i}^{(1)} \nonumber
\eeq
 
\textit{Step 1. Characteristic function and joint normality of the leaves} We first compute the characteristic function of the leaves and show how conditional independence (CI) statements enable its calculation. This will show that the leaves are jointly normal. Let $\textbf{t} \in \mathbb{R}^N$, $N=k^m$ and $\textbf{X}_{(m)}$ the vector made of the $X_{i}^{(m)}$'s. Then:
\beq
\Phi_{\textbf{X}_{(m)}}(\textbf{t}) \equiv \E[\exp(i \mathbf{t}'\mathbf{X}_{(m)})] \nonumber
\eeq
We note $X_{u,i}^{(m)}$ the leaves $X_{u}^{(m)}$ such that $X_{u}^{(m)} \leq Y_{i}$ for $i \in [1,k]$, and we note $\mathcal{L}_{i}=\{X_{u}^{(m)} \leq Y_{i}\}$ these sets of leaves. Then the scalar product $\mathbf{t}'\mathbf{X}_{(m)}$ is naturally split in $k$ sums according to these sets $\mathcal{L}_i$:
\beq
\mathbf{t}'\mathbf{X}_{(m)} =  \displaystyle\sum_{i=1}^{i=k} \mathbf{t}_i'\mathbf{X}_i^{(m)} \nonumber
\eeq
where the notation is self-explanatory. Then
\beq
\Phi_{\textbf{X}_{(m)}}(\textbf{t}) = \E\left[ \prod_{i=1}^{i=k}\exp(i \mathbf{t}_i'\mathbf{X}_i^{(m)})\right] \nonumber
\eeq
that we now condition on all the roots of the subtrees using the law of total expectation: 
\beq
\Phi_{\textbf{X}_{(m)}}(\textbf{t}) = \E\left[ \E\left[\prod_{i=1}^{i=k}\exp(i \mathbf{t}_i'\mathbf{X}_i^{(m)})| \mathbf{Y}^{(1)}\right]\right] \nonumber
\eeq
By virtue of conditional independence, we however have $\forall (u,v) \in  \mathcal{L}_{i}\times \mathcal{L}_{j}$, the property: $X_{u,i}^{(m)} \independent X_{v,j}^{(m)}|Y_{i}$ and therefore, $\forall \mathbf{t}', \forall i\neq j$, $\mathbf{t}'_i \mathbf{X}_{i}^{(m)} \independent\mathbf{t}'_j \mathbf{X}_j^{(m)}|Y_{i}$ and finally 
\beq
\forall \mathbf{t}', \mathbf{t}'_i \mathbf{X}_{i}^{(m)} \independent\mathbf{t}'_j \mathbf{X}_j^{(m)}|\mathbf{Y} \nonumber
\eeq
This enables to factorize the characteristic function as\footnote{If $X \independent Y | Z$ then $\E(XY|Z)=\E(X|Z)\E(Y|Z)$.}
\beq
\Phi_{\textbf{X}_{(m)}}(\textbf{t}) = \E\left[ \prod_{i=1}^{i=k} \E\left[\exp(i \mathbf{t}_i'\mathbf{X}_i^{(m)})| \mathbf{Y}\right]\right] \nonumber
\eeq
Others conditional independence statements of the form 
\beq
X_{u,i}^{(m)}\independent Y_{j}|Y_{i} \Rightarrow \mathbf{t}'_i \mathbf{X}_{i}^{(m)}\independent \mathbf{Y}_{-i}|Y_{i} \nonumber
\eeq
where $\mathbf{Y}_{-i}$ is the vector $\mathbf{Y}$ with the $i^{\text{th}}$ element removed, enable ones to remove redundant conditioning in the expression for the characteristic function, and one finally gets:
 \beq
\Phi_{\textbf{X}_{(m)}}(\textbf{t}) = \E\left[ \prod_{i=1}^{i=k} \E\left[\exp(i \mathbf{t}_i'\mathbf{X}_i^{(m)})| Y_i\right]\right]
\eeq
which is now calculable, because $ \E\left[\exp(i \mathbf{t}_i'\mathbf{X}_i^{(m)})| Y_i\right]$ is a known random variable thanks to recurrence hypothesis (it indeed involves only the properties of the subtrees, which are assumed by recurrence to be jointly normal), whereas the second expectation is between (products of) random variables related to the $Y_i$ only, and which are, by hypothesis, univariate normals aggregated together via a Gaussian copula. We have
\beq
\label{characintegral}
\Phi_{\textbf{X}_{(m)}}(\textbf{t}) =\int  f_\mathbf{Y}(\mathbf{y}) \prod_{i=1}^{i=k} \left( \int  d\mathbf{x}_i \exp(i \mathbf{t}_i' \mathbf{x}_i) f\left(\mathbf{x}_i|Y_i=y_i\right) \right) d\mathbf{y} 
\eeq
where $\mathbf{y}=(y_1, \ldots y_k)$,  $f_\mathbf{Y}(\mathbf{y})$ is the joint density between the random variables $Y_i$, and where  $f\left(\mathbf{x}_i|Y_i=y_i\right)$ is a conditional density function. 

The leaves $ X_j^{(m)}$ for $j \in \mathcal{L}_i$ are jointly normal by recurrence hypothesis and $Y_i=\displaystyle\sum_{j \in \mathcal{L}_i} X_j^{(m)}$. As shown in Corollary \ref{CoroCondition}, this implies that $\mathbf{X}_i^{(m)}|Y_i=y_i$ is multivariate normal and  $ f\left(\mathbf{x}_i|Y_i=y_i\right)$ is its density. As a consequence, the multiple integral over $\mathbf{x}_i$ leads to a conditional characteristic function of a multivariate normal:
\beq
\int  d\mathbf{x}_i \exp(i \mathbf{t}_i' \mathbf{x}_i) f\left(\mathbf{x}_i|Y_i=y_i\right)
\stackrel{\exists}{=} \exp\left(i \mathbf{t}'_i \tilde{\boldsymbol\mu_i} - \frac{1}{2}  \mathbf{t}'_i \tilde{\Sigma}_i  \mathbf{t}_i \right) \nonumber
\eeq
where  $\tilde{\boldsymbol\mu_i}$, as shown in \ref{CoroCondition} is affine in $y_i$ whereas the covariance matrix $\tilde{\Sigma}_i $ does not depends on $y_i$. In fact, because the means of the leaves are zero, $\tilde{\boldsymbol\mu_i}$ is even linear in $y_i$: there exists a column vector $\mathbf{B}_i$ such that $\tilde{\boldsymbol\mu_i}= \mathbf{B}_i y_i$.

Replacing in Eq.~(\ref{characintegral}), and factorizing, we get, defining $s_i= \mathbf{t}'_i \mathbf{B}_i$ and $\mathbf{s}=(s_1,\ldots,s_k)'$:
\begin{eqnarray}
\label{characfinal}
\Phi_{\textbf{X}_{(m)}}(\textbf{t}) &=&  \left(\prod_{i=1}^{i=k} \exp \left(- \frac{1}{2}  \mathbf{t}'_i \Sigma_i  \mathbf{t}_i \right) \right). \int  \prod_{i=1}^{i=k} dy_i  f_\mathbf{Y}(\mathbf{y}) \exp\left(i \mathbf{t}'_i \mathbf{B}_i y_i) \right)\nonumber \\
&=& \exp \left(- \frac{1}{2} \displaystyle\sum_{i=1}^{i=k} \mathbf{t}'_i \Sigma_i  \mathbf{t}_i \right) \int d\mathbf{y}  f_\mathbf{Y}(\mathbf{y}) \exp\left(i \mathbf{s}' \mathbf{y}) \right)\nonumber \\ 
&=& \exp \left(- \frac{1}{2} \displaystyle\sum_{i=1}^{i=k} \mathbf{t}'_i \Sigma_i  \mathbf{t}_i \right) \exp \left( -\frac{1}{2} \mathbf{s}' \tilde{\Sigma} \mathbf{s} \right) \nonumber \\
&=& \exp \left(- \frac{1}{2} \displaystyle\sum_{i=1}^{i=k} \mathbf{t}'_i \Sigma_i  \mathbf{t}_i -\frac{1}{2} \mathbf{s}' \tilde{\Sigma} \mathbf{s} \right) \nonumber \\
&\stackrel{\exists}{=} &  \exp \left(- \frac{1}{2}\mathbf{t}' A \mathbf{t} \right) \nonumber \\
\end{eqnarray}
Where we used that the integral in the second line is the characteristic function of the jointly normal distribution of $\mathbf{Y}$, and therefore there exists some matrix $\tilde{\Sigma}$ such that the integral take the form in the third line (the means of the $Y_i$ are zero), and where, in the last line, the existence of a \textit{constant} matrix $A$ is readily proven, noting that 
\beq
A_{ij}= \partial_{t_i} \partial_{t_j} \left( - \ln \Phi_{\textbf{X}_{(m)}}(\textbf{t}) \right) \nonumber
\eeq
and that (see the fourth line), the form is not more than quadratic in $t_i$'s because the $s_i$'s are linear combinations of the $t_j$'s. 

This proves that the existence of a matrix $A$ such that the joint characteristic function is an exponential of the quadratic form defined by $A$. Moreover, this particular form of the characteristic function implies that $A$ is also the covariance matrix of $\mathbf{X}$, and therefore is non-negative definite. As it is well-known, a characteristic function for $\mathbf{X}$ of the form above with a non-negative definite matrix $A$ implies  that $ \mathbf{X}$ is jointly normal. $\Box$

\textit{Step 2. Explicit calculation of the covariance matrix}
The matrix $A$ could be computed directly by an explicit evaluation of Eq.~(\ref{characfinal}) above. There is however an easier route to do so by exploiting the many symmetries there are in the problem.

The block diagonal terms of the covariance matrix are easily derived. Consider the leaves $X_{u,i}^{(m)}$ that are descendants of $Y_{i}$ for a given $i$. By recurrence hypothesis, these leaves are jointly normal with a covariance matrix $\mathcal{C}^{(m-1)}$ as given in the theorem, except the necessary change in notations explained at the beginning of the proof. This holds for every $i$, hence showing that the covariance matrix $\mathcal{C}^{(m)}$ must have the block-diagonal structure given in the theorem. 

Off-diagonal block terms: For any two pair $i \neq j \in [1,k]^2$ we compute the covariance matrix elements\footnote{To make some contact with the step 1 above, the off-diagonal elements clearly come from the linear relation between $s$ and $t$, see Eq.(\ref{characfinal})} 
$\text{cov}(X_{u,i}^{(m)}, X_{v,j}^{(m)})$ for $(u,v) \in \mathcal{L}_{i}\times \mathcal{L}_{j}$. These elements of the covariance matrix corresponds to the elements found inside the $i,j$ block of the covariance matrix when written in $k\times k$ blocks of size $k^{m-1}$. Applying two times the conditional independence as done in step 1, we have \footnote{I.e. using first the total law of expectation, and conditioning on the roots of the subtrees in question, applying afterwards the conditional independence hypothesis with e.g. $X_{u,i}^{(m)} \independent X_{v,j}^{(m)}|Y_{i}$. Finally we apply another conditional independence statement: $X_{u,i}^{(m)}\independent Y_{j}|Y_{i}$ implying that $\E[X_{u,i}^{(m)}|Y_{i},Y_{j} ]=\E[X_{u,i}^{(m)}|Y_{i}]$.}, using that the means vanish:
\begin{eqnarray}
\text{cov}(X_{u,i}^{(m)}, X_{v,j}^{(m)}) &=& \E[X_{u}^{(m)}X_{v}^{(m)}] \nonumber \\
 &=& \E[\E[X_{u,i}^{(m)}X_{v,j}^{(m)}|Y_{i},Y_{j}]] \nonumber \\
 &=& \E[\E[X_{u,i}^{(m)}|Y_{i},Y_{j} ] \E[X_{v,j}^{(m)}|Y_{i},Y_{j} ]] \nonumber \\
 &=& \E[\E[X_{u,i}^{(m)}|Y_{i} ] \E[X_{v,j}^{(m)}|Y_{j} ]] \nonumber \\
\end{eqnarray}
This shows again how CI enables to compute these off-diagonal terms of the covariance matrix. Indeed $\E[X_{u}^{(m)}|Y_{i}]$ depends only on the subtree, whereas the expectation of the product of expectations in last line only depends on the last aggregation step from level $(1)$ to the root.

More precisely, $\E[X_{u}^{(m)}|Y_{i}]$ could be explicitly computed using similar techniques as the one found in Corollary \ref{CoroCondition} and using the recurrence hypothesis. It is however not necessary. Indeed it is enough to notice that, because the leaves are all identical and because the exchangeable (ie. with equicorrelation matrix) copula is used, $\E[X_{u}^{(m)}|Y_{i}]$ cannot depend on $u$ nor on $i$.

Similarly the exchangeable copula used at last aggregation step does not introduce any asymmetry between the roots of the subtrees $Y_i$ and $Y_j$, and therefore $\text{cov}(X_{u,i}^{(m)}, X_{v,j}^{(m)})$ must be a constant for all $i \neq j$ and for all $u,v$. We denote the off-block-diagonal elements of $\mathcal{C}^{(m)}$ by 
\beq
\text{cov}(X_{u,i}^{(m)}, X_{v,j\neq i}^{(m)}) = \sigma_{(m)}^2 \beta_{m-1} \nonumber
\eeq
and any $i \neq j$ off-diagonal blocks by
\beq
 B_{(m-1)}= \beta_{m-1} J_{k^{m-1}} \nonumber
\eeq
The last task is to compute $\beta_{m-1}$. First we note that $X_{u,i}^{(m)}|Y_{i}=y_i$ is a Gaussian of mean $y_i/N'$, where $N'=k^{m-1}$ is the number of leaves in the subtree $Y_i$. This is clear by symmetry, but can also be proven\footnote{Along the lines of theorem \ref{theocondition} and corollary \ref{CoroCondition}. Choose one particular subtree, e.g. the one issued from $Y_1$. Its covariance matrix is known by recurrence hypothesis. Construct the bivariate normal $(X_1,Y_1)$ and then condition on $Y_1=y_1$. Evaluate formula for the mean of $X_1|Y_1=y_1$, and find $\tilde{\mu}=\gamma y$ where $\gamma$ is the ratio between the sum of elements of the first line of the covariance matrix of the subtree, and the sum of all its elements. By symmetry of the covariance matrix given by recurrence, this ratio must be $1/N'$.} directly.
Then we have that 
 \begin{equation}
\text{cov}(X_{u,i}^{(m)}, X_{v,j\neq i}^{(m)}) = \sigma_{(m)}^2 \beta_{m-1} =\frac{k^2}{N^2}\E[Y_i Y_j] \nonumber
\end{equation}
where, as already argued, $\E[Y_i Y_j]$ is independent of the pair $i\neq j$ because the $Y_i$ have all the same marginals and are aggregated via an exchangeable copula. Since  $Z^2= \sum Y_i^2+\sum_i\sum_{j\neq i}Y_i Y_j$, and noting that there are $k(k-1)$ terms in the double sum, we get
\beq
\E[Y_i Y_j]= \frac{1}{k(k-1)}\left(\E[Z^2]-k \E[Y_i^2]\right)\nonumber
\eeq
Because the means vanish, we note that $\E[Y_i^2]=\sigma_{(1)}^2$ and $\E[Z^2]=\sigma_Z^2$. The recurrence hypothesis gives the relation between the variances $\sigma_{(1)}^2= \sigma_{(m)}^2 (k+(k^2-k)\rho)^{m-1}$, and we need to prove that it also holds at the upper level, ie. that $\sigma_Z^2=\sigma_{(1)}^2 (k+(k^2 -k)\rho)$. This is straightforward from the fact that $Y_i \sim \mathcal{N}(0,\sigma_{(1)}^2)$ and the aggregation mechanism, and follows as a special case of theorem \ref{Maintheo}.
Therefore we get, all calculations done:
\begin{eqnarray}
\beta_{m-1} &=& \frac{1}{\sigma_{(m)}^2 }\frac{k^2}{N^2} \frac{1}{k(k-1)}\left(\E[Z^2]-k \E[Y_i^2]\right) \nonumber \\
&=&\rho \frac{k^2}{N^2}  \frac{\sigma_{(1)}^2}{\sigma_{(m)}^2 } \nonumber \\
&=&  \rho \left( \frac{1}{k} +\left( 1-\frac{1}{k}\right) \rho\right)^{m-1} \nonumber
\end{eqnarray}
thus establishing the claim. $\Box$

\textit{Step 3: Exhausting the conditional independence statements} 
We showed how to use \textit{some} of the conditional independence (CI) statements to prove that the leaves are jointly normal with a covariance matrix completely determined and given as in the theorem. However the claim is also that any other CI statements are then satisfied. This is what we show now. 

To start with we note that  because of the aggregation mechanism, any node in the $T_{k,m}$ tree is either a leave or a sum of leaves. In any case, it can be written as a sum of leaves over a given set of indices:
\beq
\forall W \in \mathcal{T}_{k,m}, \exists J_W: W= \displaystyle\sum_{i \in J_W} X^{(m)}_i \nonumber
\eeq
As a consequence, any CI statements given in definition \ref{CItree} can be deduced from the following ones (e.g. by adding them together, etc):
\beq
\forall W \in \tilde{T}, \forall i \in J_W, \forall j \notin J_W:  X_i^{(m)}\independent X_j^{(m)} |W
\eeq
where $\tilde{T}$ is the set of nodes that are not leaves of the tree nor the root.

This set of relations are then expressed as algebraic equations between the coefficients of the covariance matrix $\mathcal{C}^{(m)} $. To this end one follows the techniques found in theorem \ref{theocondition} and corollary \ref{CoroCondition}, and construct a new multivariate normal:
\beq
\chi  =  (X_i^{(m)}, X_j^{(m)}, W)= \left( X_i^{(m)}, X_j^{(m)}, \sum_{u \in J_W} X_u^{(m)}\right) \stackrel{\exists \mathbf{M}}{=} \mathbf{M}.\mathbf{X}^{(m)} \nonumber
\eeq
where $\mathbf{M}$ is a $3\times N$ matrix with first line equals to $\delta_{ik}$ with running $k$, second line equals to $\delta_{jk}$ with running $k$, and third line is full of 1's for indices in $J_W$, zero otherwise. Then $\chi$ is multivariate normal with covariance matrix given by $\Sigma=\mathbf{M} \mathcal{C}^{(m)} \mathbf{M}'$. This matrix is easily computed and reads
\[
\Sigma = 
 \begin{pmatrix}
  \mathcal{C}^{(m)}_{ii} & \mathcal{C}^{(m)}_{ij} &\mathcal{C}^{(m)}_{iJ_W}  \\
   \mathcal{C}^{(m)}_{ij} & \mathcal{C}^{(m)}_{jj} &\mathcal{C}^{(m)}_{jJ_W}  \\
  \mathcal{C}^{(m)}_{iJ_W} & \mathcal{C}^{(m)}_{j J_W} &\mathcal{C}^{(m)}_{J_WJ_W}  
 \end{pmatrix}
\]
where $  \mathcal{C}^{(m)}_{iJ_W} \equiv \sum_{u \in J_W}  \mathcal{C}^{(m)}_{iu} $ and  $  \mathcal{C}^{(m)}_{J_W J_W} \equiv \sum_{u,v \in J_W^2}  \mathcal{C}^{(m)}_{uv} $. Then, conditioning on $W$, we have that $X_i,X_j|W$ is normal with a $2\times2$ covariant matrix $\tilde{\Sigma}$ given by 
\[
\tilde{\Sigma} = 
 \begin{pmatrix}
  \mathcal{C}^{(m)}_{ii} & \mathcal{C}^{(m)}_{ij} \\
   \mathcal{C}^{(m)}_{ij} & \mathcal{C}^{(m)}_{jj} 
 \end{pmatrix}
- \frac{1}{\mathcal{C}^{(m)}_{J_WJ_W}}
 \begin{pmatrix}
 \mathcal{C}^{(m)}_{iJ_W} \mathcal{C}^{(m)}_{iJ_W}& \mathcal{C}^{(m)}_{iJ_W}\mathcal{C}^{(m)}_{jJ_W} \\
  \mathcal{C}^{(m)}_{iJ_W}\mathcal{C}^{(m)}_{jJ_W} &\mathcal{C}^{(m)}_{jJ_W} \mathcal{C}^{(m)}_{jJ_W}
 \end{pmatrix}
\]
The conditional independence requires the off-diagonal terms to vanish. Therefore we conclude that all CI statements that can be made in the Gaussian $T_{k,m}$ tree with covariance matrix $\mathcal{C}^{(m)}$ are of the following form:
\beq
\label{CIcond}
\forall W \in \tilde{T}, \forall i \in J_W, \forall j \notin J_W, \mathcal{C}^{(m)}_{ij}=\frac{ \mathcal{C}^{(m)}_{iJ_W}\mathcal{C}^{(m)}_{jJ_W}}{\mathcal{C}^{(m)}_{J_WJ_W}}
\eeq
We now prove that these conditions hold. For this we compute each of these terms separately. We first prove proposition \ref{proposeff} on effective couplings that we shall use below. Two nodes $X_i^{(m)}$ and $X_j^{(m)}$ connected via a  node at level  $p$ but not connected by any node at level $p'>p$ are leaves of a subtree $\mathcal{T}_{k,m-p}$ for which the covariance matrix is known by strong recurrence hypothesis (and if $p=0$, the covariance matrix has been derived in the previous steps). Moreover, as the node at level $p$ is the first node connecting these two leaves, $\textrm{cov}(X_i^{(m)},X_j^{(m)})$ is given by the off-block diagonal elements of $\mathcal{C}^{(m-p)}$, hence by $\beta_{m-p-1}$. This establishes the claim.  

Now, for any $W$ there exists $r$,$p$ such that $W=X_r^{(p)}$ with $p \neq 0$ and $p\neq m$. Thus $W$ is the root of a $\mathcal{T}_{k,m-p}$ Gaussian tree, and by virtue of corollary \ref{corosum},  $ \mathcal{C}^{(m)}_{J_W J_W} \equiv \sum_{u,v \in J_W^2}  \mathcal{C}^{(m)}_{uv} $ equals the variance of $W$, which, by virtue of theorem \ref{Maintheo} reads $\sigma_{(p)}^2$. Hence
\beq
\mathcal{C}^{(m)}_{J_WJ_W}= \sigma_{(p)}^2 \nonumber
\eeq
To compute $\mathcal{C}^{(m)}_{iJ_{W}} \equiv \sum_{u \in J_W}  \mathcal{C}^{(m)}_{iu}$ with $i \in J_W$ we recognize that the total number of terms in the sum reads $k^{m-p}$, which can be split as $k^{m-p}=(k^{m-p}-k^{m-p-1}) + (k^{m-p-1}-k^{m-p-2})+ \ldots + (k-1) +1$ which corresponds to the numbers of leaves having a resp. effective coupling with the leave $i$ given by $ \beta_{m-s-1}$ with a running $s$, i.e.
\begin{eqnarray}
\mathcal{C}^{(m)}_{iJ_{W}}/\sigma_{(m)}^2 &=&1+\displaystyle\sum_{s=p}^{s=m-1} (k^{m-s}-k^{m-s-1}) \beta_{m-s-1} \nonumber \\
&=& 1+ \displaystyle\sum_{s=0}^{s=m-p-1} \rho (k-1) \left(1+(k-1)\rho\right)^{s} \nonumber \\
&=&(1+(k-1)\rho)^{m-p} \nonumber
\end{eqnarray}
Thus,
\beq
\mathcal{C}^{(m)}_{iJ_{Y_W}}=\sigma_{(m)}^2 (1+(k-1)\rho)^{m-p} \nonumber
\eeq

For $j\notin J_W$, $\mathcal{C}^{(m)}_{jJ_W}= \sum_{u \in J_W} \mathcal{C}^{(m)}_{ju}$ has again $k^{m-p}$ terms, which are this time all equal to each other, but this constant depends on the position of the leave $j$ with respect to the set $J_W$, i.e. at which level $l$ one finds the first node connecting these leaves (with of course $l>p$). Then we have (using again the effective correlation coefficients):
\beq
\mathcal{C}^{(m)}_{jJ_W}= \sigma_{(m)}^2 k^{m-p}\rho \left( \frac{1}{k} + \left( 1 -\frac{1}{k}\right) \rho\right)^{m-l-1} \nonumber
\eeq
Finally $ \mathcal{C}^{(m)}_{ij}$ is computed as is computed $\mathcal{C}^{(m)}_{jJ_W}$ but without any summation in this case. Hence: 
\beq \mathcal{C}^{(m)}_{ij}=\sigma_{(m)}^2 \rho \left( \frac{1}{k} + \left( 1 -\frac{1}{k}\right) \rho\right)^{m-l-1} \nonumber
\eeq
 
We are now in position to form the right hand side of Eq.~(\ref{CIcond}). For all $(i,j)$, i.e for all $(p, l)$, the r.h.s. reads
\begin{eqnarray}
\frac{ \mathcal{C}^{(m)}_{iJ_W}\mathcal{C}^{(m)}_{jJ_W}}{\mathcal{C}^{(m)}_{J_WJ_W}}&=&\frac{\sigma_{(m)}^4  (1+(k-1)\rho)^{m-p} k^{m-p}\rho \left( \frac{1}{k} + \left( 1 -\frac{1}{k}\right) \rho\right)^{m-l-1}}{ \sigma_{(p)}^2} \nonumber \\
&=&\frac{\sigma_{(m)}^2 (k+(k^2-k)\rho)^{m-p} \rho \left( \frac{1}{k} + \left( 1 -\frac{1}{k}\right) \rho\right)^{m-l-1}}{  (k+(k^2-k)\rho)^{m-p} } \nonumber \\
&=& \sigma_{(m)}^2 \rho \left( \frac{1}{k} + \left( 1 -\frac{1}{k}\right) \rho\right)^{m-l-1}\nonumber \\
&=& \mathcal{C}^{(m)}_{ij} \nonumber
\end{eqnarray}
Where the relation between variances has been used: $\sigma_{(p)}^2=\sigma_{(m)}^2(k+(k^2-k)\rho)^{m-p}$
\begin{flushright}
$\Box$
\end{flushright}
%

\end{document}